\documentclass[]{spie}  

\usepackage[dvipsnames]{xcolor}
\newcommand{\bad}[1]{\textcolor{BrickRed}{#1}}
\newcommand{\nom}[1]{\textcolor{YellowOrange}{#1}}
\newcommand{\good}[1]{\textcolor{PineGreen}{#1}}

\usepackage{amsmath,amsfonts,amssymb}
\usepackage{graphicx}
\usepackage{multirow}
\usepackage[colorlinks=true, allcolors=blue]{hyperref}
\title{Exoplanet Imaging Data Challenge, phase II: Comparison of algorithms in terms of characterization capabilities}
\author[a]{Cantalloube~F.}
\author[b]{Christiaens~V.}
\author[c]{Cantero~C.}
\author[d]{Cioppa~A.}
\author[e]{Nasedkin~E.}
\author[b]{Absil~O.}
\author[a]{Delorme~P.}
\author[f,g]{Wang~J.-J.}
\author[h]{Bonse~J.~M.}
\author[i]{Daglayan~H.}
\author[b]{Dahlqvist~C.-H..}
\author[b]{Guyot~N.} 
\author[b]{Juillard~S.}
\author[j]{Mazoyer~J.}
\author[e]{Samland~M.}
\author[b]{Sabalbal~M.}
\author[k]{Ruffio~J.-B.}
\author[b]{Van~Droogenbroeck~M.}

\affil[a]{Univ. Grenoble Alpes, CNRS, IPAG, F-38000 Grenoble, France}
\affil[b]{Space Sciences, Technologies \& Astrophysics Research (STAR) Institute, Universit\'e de Li\`ege, All\'ee du Six Ao\^{u}t 19c, B-4000 Li\`ege, Belgium}
\affil[c]{Geneva Observatory, University of Geneva, Chemin Pegasi 51, 1290 Versoix, Switzerland}
\affil[d]{Montefiore Institute, Universit\'e de Li\`ege, 4000 Li\`ege, Belgium}
\affil[e]{Max-Planck-Institut f\"{u}r Astronomie, K\"{o}nigstuhl 17, Heidelberg 69117, Germany}
\affil[f]{Center for Interdisciplinary Exploration and Research in Astrophysics (CIERA), Northwestern University, 1800 Sherman Ave, Evanston, IL, 60201, USA}
\affil[g]{Department of Physics and Astronomy, Northwestern University, 2145 Sheridan Rd, Evanston, IL 60208, USA}
\affil[h]{Institute for Particle Physics and Astrophysics, ETH Zurich, Wolfgang-Pauli-Strasse 27, 8093, Zurich, Switzerland}
\affil[i]{ICTEAM Institute, UCLouvain, Louvain-la-Neuve, Belgium}
\affil[j]{LESIA, Observatoire de Paris, Universit\'e PSL, CNRS, Sorbonne Universit\'e, Universit\'e de Paris, 5 place Jules Janssen, F-92195 Meudon, France}
\affil[k]{Department of Astronomy \& Astrophysics, University of California, San Diego, La Jolla, CA 92093, USA}

\authorinfo{Further author information: (Send correspondence to F.C.)\\ 
F.C.: E-mail: faustine.cantalloube@univ-grenoble-alpes.fr}

\begin{document} 
\maketitle

\begin{abstract}
In this communication, we report on the results of the second phase of the \emph{Exoplanet Imaging Data Challenge} started in 2019. This second phase focuses on the characterization of point sources (exoplanet signals) within multispectral high-contrast images from ground-based telescopes. We collected eight data sets from two high-contrast integral field spectrographs (namely Gemini-S/GPI and VLT/SPHERE-IFS) that we calibrated homogeneously and in which we injected a handful of synthetic planetary signals (ground truth) to be characterized by the data challenge participants. 
The tasks of the participants consist of (1) extracting the precise astrometry of each injected planetary signals, and (2) extracting the precise spectro-photometry of each injected planetary signal. Additionally, the participants may provide the 1-sigma uncertainties on their estimation for further analyses. When available, the participants can also provide the posterior distribution used to estimate the position/spectrum and uncertainties. 
The data are permanently available on a Zenodo repository and the participants can submit their results through the EvalAI platform. 
The EvalAI submission platform opened on April 2022 and closed on the 31st of May 2024. 
In total, we received $4$ valid submissions for the astrometry estimation and $4$ valid submissions for the spectrophotometry (each submission, corresponding to one pipeline, has been submitted by a unique participant). 
In this communication, we present an analysis and interpretation of the results. 
\end{abstract}


\section{INTRODUCTION}
\label{sec:intro}  
The \emph{Exoplanet Imaging Data Challenge} is an initiative led by several members of the exoplanet high-contrast imaging community. 
The goal of this initiative is to benchmark and compare the various existing post-processing techniques that are dedicated to exoplanet direct detection via high-contrast imaging. Indeed, the last ten years have seen the implementation of a number of advanced image processing techniques whose goal is to further carve out the starlight residuals in the images so as to reach the highest contrast possible, especially at close angular separation from the star (where most planets are expected), towards detecting faint planetary signals. 

Current high-contrast images delivered by state-of-the-art instruments are affected by bright starlight residuals hindering the presence of exoplanets. These residuals come either from adaptive optics residuals\cite{cantalloube2020wdh} (usually short-lived so showing as smooth structures), low order residuals such as jitter and low wind effect\cite{sauvage2016} (usually varying fast and affecting the inner region), non-common path aberrations\cite{vigan2022} (fixed aberrations to calibrate, internal turbulence showing smooth structures and quasi-static aberrations), and any other error related to the detection procedure (e.~g.~bad pixels, miscalibration). The ultimate limit is given by the photon noise of the target star residuals. 
Usually, before applying dedicated post-processing technique to gain one or two orders of magnitude in contrast, we apply some \emph{pre-processing} (also called cosmetic, or calibration) including the dark subtraction, flat fielding, distortion correction, wavelength calibration, centering\footnote{See e.g. the review talk summarizing the importance of these steps \url{https://www.youtube.com/watch?v=0v4q6VN4b9M}.
} etc. 
The task of post-processing techniques is to deal with the effect of all the residual optical aberrations, which distribute starlight across the field of view on different spatial and temporal scales\cite{cantalloube2019}, in order to identify and characterize circumstellar signals.

The \emph{Exoplanet Imaging Data Challenge} (EIDC) is organized in several phases, and is evolving to include more features and adapt to the latest findings. 
The first phase\cite{Cantalloube2020eidc} (2020), was focused on the detection of planetary signals in the field of view. 
This second phase is focused on the characterization of planetary signals. 
Upcoming phases will include extended features, the use of high resolving power spectroscopy, the exploitation of large data base and/or multi-epoch, telemetry and/or environmental data, and simulated images for the Extremely Large Telescope suites of instrument. 
The EIDC working group plans to publish its work and various results regularly, in particular every two years in the form of an \textit{SPIE Astronomical Telescope + Instrumentation proceeding}. 

Data from the various phases of the \emph{Exoplanet Imaging Data Challenge} are permanently available on a Zenodo repository, so that anyone can use the ensemble of curated data to test and compare the capabilities of different algorithms. We are also providing analysis tools, including the metrics and comparison set of the corresponding phase. The \emph{Exoplanet Imaging Data Challenge} intends to make the process of testing new algorithms more straightforward and robust, via the use of standard metrics and benchmark data set offered to the community. 
People are more than welcome to use the various resources from the \emph{Exoplanet Imaging Data Challenge} platform for test and scientific publication (with due acknowledgement).

The data offered for this phase of the data challenge as well as the metrics used for the leaderboard are previously described \cite{eidc2022} in a 2022 SPIE proceeding. 
More information about this community-led initiative can be found on the dedicated website\footnote{\url{https://exoplanet-imaging-challenge.github.io/}} (context, data, submission procedure, results and extended bibliography).

\section{Exoplanet characterization with high-contrast imaging}
Direct imaging of exoplanets, among other techniques, offers a unique window on planetary systems: (1) first, it provides a full picture of the system to study its architecture (interaction with circumstellar disks, multiplicity etc.); (2) second, the brightness estimation of the substellar companions provide with estimates of planet properties, such as its mass (from the mass luminosity relationships, linked to evolutionary track models), and if multi-wavelength images are additionally offered, spectral forward models and spectra retrievals unveil information on the atmospheric structure and composition; (3) precise astrometric follow-up may provide orbital constraints that can help us understanding the formation and evolution processes at stake such as migration, scattering and resonances, as well as dynamical mass estimate. 
The target dataset detection limit is computed and used to set statistical constraints on the occurrence rate of exoplanets of a given mass, at a given separation, around this type of target star. 
With current instruments, direct imaging of exoplanets is sensitive to the population of young giant gaseous planets at relatively large orbit that appear to be quite rare\cite{nielsen2019gpies,vigan2021shine}.

In terms of image processing, the characterization of a planetary signal detection consists in extracting two observables: (I) the \textit{astrometry}, which is the  estimated position of the planetary companion(s) in the image, and (II) the \textit{spectro-photometry}, which is the  estimated luminosity in terms of contrast of the planetary companion(s) in the images. 
From these two estimates, observers can analyze and interpret the physical nature of the planet and, on larger scales, infer the formation and evolution processes of planetary systems. Obtaining accurate and precise observable is therefore of acute importance in the science of exoplanets.

\subsection{Reminder on ADI post-processing techniques}
As a reminder the most common method used for processing high-contrast images is based on differential imaging (DI), in particular making use of the angular diversity\cite{Marois2006} offered by pupil-tracking observations from telescopes having an alt-az mount. Indeed, in such a configuration, the orientation of the telescope pupil (where most optical aberrations arise) is kept along a given direction in the image cube, whereas the orientation of the field of view in the focal plane rotates with the parallactic angles as the Earth rotates during the observation sequence. In the resulting temporal cubes of images, the starlight residuals (nuisance term, also called stellar glare) is kept more or less stable in the field of view, while the off-axis astrophysical objects rotate around the image center (where the target star is located) with the parallactic angle during the observation. 

Differential Imaging can be summarized in four steps: (1) estimating the starlight distribution in the field of view, also called \emph{reference PSF} or \emph{model PSF}; (2) Subtracting it to the science image(s), also called \emph{PSF subtraction}; (3) repeat the subtraction procedure for each frame of the image cube then rotate the frames in order to align the circumstellar signals and combine them to obtain a \emph{residual map}; (4) build a \emph{detection map} (e.g. an SNR map) giving the probability of detecting a signal in the image field of view\footnote{For a review of post-processing technique and concepts of ADI, please visite our website \url{https://exoplanet-imaging-challenge.github.io/biblio/}.}. The essential differences between current DI-based techniques are focused on the  step (1), critical to obtaining an effective stellar glare subtraction. 
For instance, classical ADI\cite{Marois2006} consists in using the temporal median of the cube as the reference PSF. 
Another widely used method is LOCI\cite{lafreniere2007loci}, which consists in finding the linear combination of images that minimizes the variance of the residuals between the reference PSF and the science image. At last, another mainstream method is to make a principal component analysis (PCA) of the data cube to build the reference PSF as the sum of the first few principal components.

\subsection{Characterization methods with ADI post-processing techniques}
As of today, there are mainly two ways of characterizing exoplanets detected within high-contrast images processed with ADI-based techniques: (i) the injection of NEGative Fake Companion\cite{marois2010negfc,lagrange2010negfc} (NEGFC) in the data set, and (ii) the Forward Modeling (FM) approaches\cite{Mugnier2009,Pueyo2016}. 

The NEGFC technique consists in injecting a synthetic planetary signal (i.e.~the non-coronagraphic / non-saturated PSF of the instrument, usually acquired during calibration before or after the coronagraphic / saturated image cube) with negative intensity in the raw image cube, at the rough position of the detection and with a rough flux estimate of the detection. The image cube with the negative injection is then processed with exactly the same user-parameters as for the planetary detection (e.g.~with PCA, using exactly the same number of principal components to build the PSF model to be subtracted to the images). 
Using various methods (such as Nelder-Mead simplex algorithm, MCMC sampler or nested-sampling), this procedure is repeated with various positions and fluxes (around the first guess) for the negative injection, until the residuals are minimized at the location of the planet, providing with the astrometry (position of the negative injection that minimizes the residuals) and the photometry (intensity of the negative injection that minimizes the residuals). 

The FM technique usually models the planetary signature (again as the non-coronagraphic / non-saturated PSF of the instrument) and tracks it in the data within an inverse problem framework. 
At the exception of the PACO algorithm\cite{flasseur2018paco}, which does not perform any subtraction, the FM technique is done after step (2). It means that under a certain assumption of the residual noise distribution in the residual frames (2) - and in the case of PACO, the raw frames at (1) - the model of the planetary signature is fitted to the actual data. By construction, as the flux of the model planetary signature is the optimized parameter, the SNR map, used as a detection map, is a by-product (defined as $\hat{\phi}/\sigma(\hat{\phi})$, with $\hat{\phi}$ the estimated flux and $\sigma(\hat{\phi})$ the uncertainties of the estimated flux).

In the previous EIDC publication\cite{eidc2022}, we compared these two basic approaches on the so-called \emph{training dataset} (dubbed \textit{sphere0}): one using a PCA-SADI procedure, as implemented in the VIP package\cite{Christiaens2022b}, followed by the NEGFC approach, and the other using ANDROMEDA\cite{cantalloube2015andro} in its classical flavor (using a L2-norm minimization because the noise distribution is assumed to be white and Gaussian). In the following, we consider the PCA-SADI+NEGFC procedure as the baseline to analyze and compare the results, which includes the uncertainties (under the assumption of a Gaussian distribution of the residuals). 
For this baseline, the uncertainties are estimated by injecting hundreds of synthetic planetary signals in the raw images at the same angular separation as the considered one, estimate their parameters, and then fitting a Gaussian to the deviation between the estimated parameters and the injected ones (ground-truth).

We received a total of $4$ valid submissions on the \emph{EvalAI} platform, from four different participants. None of the participants included neither the uncertainties nor the posterior in their results.  
The four submissions were obtained using: 
\textbf{RSM}, the Regime Switching Model\cite{dahlqvist2020rsm}, implemented with a forward modeling approach, following the description in Dahlqvist et al.\cite{dahlqvist2021rsmfm} (2021); 
\textbf{pyKLIP-FM}, the Karhunen-Loève Image Processing\cite{Soummer2012}, as implemented in the pyKLIP\cite{Wangpyklip} pipeline, using Forward Modeling for the characterization described in Pueyo et al.\cite{pueyo2016fm} (2016); 
\textbf{AMAT}, the Alternating Minimization Algorithm with Trajectory\cite{daglayan2023amat}, which uses a L1-norm low rank approximation to enhance the PCA-based PSF subtraction, associated to a forward model approach similar to ANDROMEDA but under the assumption of a Laplacian residual noise distribution (so using a L1-norm minimization to estimate the flux of the planet signature) that corresponds better to the actual distribution of the residuals after the subtraction; 
\textbf{ANDROMEDA}, in its classical\cite{cantalloube2015andro} ADI flavour, that is to say using a Gaussian residual assumption for the flux estimation and running every wavelength cube separately to combine them at the end (note that we only received the astrometry estimations from the latter but no spectrophotometry).

\section{Results: Astrometry}
As a reminder, for the astrometry estimation of the Exoplanet Imaging Data Challenge second phase, we asked the participants to provide with (1) the estimated position of each injected planetary signals in the provided data set, (2) optionally the corresponding 1$\sigma$ uncertainties of the estimations, and (3) optionally the corresponding posterior distribution used to estimate the position and its uncertainties. As for the metric used for the ranking of the submitted results, the EIDC phase 2 working group decided to compute the distance (in the sense of the L2-norm, euclidian distance) between the estimated value and the ground truth value. All the details about the provided data sets, submission process and chosen metrics for comparison and ranking can be found in our previous SPIE publication\cite{cantalloube2022eidc}. 

\subsection{Astrometry: baseline results as a function of the dataset}
The provided data sets come from two different instruments (namely GPI\cite{Macintosh2008} and SPHERE-IFS\cite{Beuzit2019}). For each instrument, we provided 4 data sets under very different observing conditions (see previous publication and website for more details about it). As a first qualitative interpretation, we compare the astrometry estimated by the baseline algorithm for each data set. The astrometry is inferred with PCA-ADI in individual channels (in the top 5 channels in terms of SNR ) at the first guess location found with PCA-SADI. The results on the GPI images are shown in Fig.~\ref{fig:data_gpi} and on the SPHERE-IFS images in Fig.~\ref{fig:data_sphere} (actual position of the injections in the images, residuals map, SNR map and astrometry error to the ground-truth). Quantitative results, including the comparison metric used, are presented in Tab.~\ref{tab:agpi} and Tab.~\ref{tab:asph} for the GPI and SPHERE-IFS data respectively.

\begin{table}[!h]
\caption[agpi] 
   {\label{tab:agpi} Results of the baseline method (PCA-ASDI+NEGFC) on the four GPI data sets, for each injection. The color of the data set ID (first column) indicates the observing conditions as good (green), medium (orange) and bad (red). See the website for more details on the data. The \emph{Injection} column corresponds to the injections as labeled in the figures all along this paper. For each injected planetary signal, $\mathrm{SNR}_{bsl}$, $\delta_{bsl}$, and $\theta_{bsl}$, are respectively the signal-to-noise ratio, the angular separation, and the position angle estimated with the baseline method. The \emph{Location} column indicates qualitatively where the injection is placed in the field of view. $D^{GT}_{astro}$ is the metric chosen for the astrometry (L2-norm distance between the estimation and the ground-truth). $D^{GT}_{photo}$ is the metric chosen for the spectrophotometry (normalized L1-norm distance between the estimation and the ground-truth.)}
\begin{center}
\begin{tabular}{|l| c | c | c | c | c | | c  || c |} 
\hline
ID & Injection & $\mathrm{SNR}_{bsl}$ & $\delta_{bsl} \, [\mathrm{mas}]$ & $\theta_{bsl} \, [\deg]$  & Location & $\mathbf{D}^{GT}_{astro}$ & $\mathbf{D}^{GT}_{photo}$\\
\hline \hline
            & b & $8.3$ & $265.58\pm 0.11 $& $94.28\pm0.033$ & Coronagraph IWA & $\mathbf{0.004}$ & $\mathbf{ 0.559 }$ \\
\good{gpi1} & c & $5.5$ & $486.75\pm3.80$ & $-154.17\pm0.17$& sitting in bright speckle & $\mathbf{0.354}$ & $\mathbf{ 0.423 }$ \\
            & d & $32.3$ & $978.50\pm0.52$ & $132.22\pm0.06$& Satellite spot & $\mathbf{0.053}$ & $\mathbf{ 0.023 }$ \\
\hline
            & b & $17.0$ & $436.66\pm 1.11 $& $26.20\pm0.06$ & AO dark-hole & $\mathbf{0.099}$ & $\mathbf{ 0.067 }$ \\
$\multirow{-2}{*}{\nom{gpi2}}$ & c & $5.2$ & $166.69\pm4.18$ & $-179.07\pm1.65$& Coronagraph IWA & $\mathbf{0.609}$ & $\mathbf{ 0.209 }$ \\
\hline
            & b & $10.0$ & $284.32\pm0.60$  & $-63.24\pm0.21$  & AO dark-hole, close & $\mathbf{0.043}$ & $\mathbf{ 0.073 }$ \\
\nom{gpi3}  & c & $4.9$  & $144.89\pm1.53$  & $160.24\pm0.92$  & Coronagraph IWA & $\mathbf{0.138}$ & $\mathbf{ 0.18 }$ \\
            & d & $47.7$ & $ 793.40\pm0.23$ & $-168.57\pm0.02$ & AO dark-hole, far & $\mathbf{0.004}$ & $\mathbf{ 0.013 }$ \\
\hline
            & b & $16.1$ & $645.06\pm0.23$  & $12.95\pm0.02$   & AO dark-hole & $\mathbf{0.024}$ & $\mathbf{ 0.212 }$ \\
$\multirow{-2}{*}{\bad{gpi4}}$ & c & $4.2$  & $151.22\pm1.40$  & $-153.84\pm0.25$  & Coronagraphic IWA & $\mathbf{0.134}$ & $\mathbf{ 3.003 }$ \\
\hline
\end{tabular}
\end{center}
\end{table}

\begin{table}[!h]
\caption[asph]{\label{tab:asph} Results of the baseline method (PCA-ASDI+NEGFC) on the four SPHERE-IFS data sets, for each injection. The \emph{sphere3} data set is taken under good observing conditions but the target star is faint. See  description of Tab.~\ref{tab:agpi}.} 
\vspace{-0.4cm}
\begin{center}
\begin{tabular}{|l| c | c | c | c | c | | c || c |} 
\hline
ID & Injection & $\mathrm{SNR}_{bsl}$ & $\delta_{bsl} \, [\mathrm{mas}]$ & $\theta_{bsl} \, [\deg]$  & Location & $\mathbf{D}^{GT}_{astro}$ & $\mathbf{D}^{GT}_{photo}$\\
\hline \hline
                & b & $4.8$ & $428.47\pm2.22$ & $56.63\pm0.09$  & AO dark-hole, far & $\mathbf{0.035}$ & $\mathbf{ 1.618 }$\\
\good{sphere1}  & c & $7.7$ & $144.34\pm0.51$ & $-18.59\pm0.34$ & Coronagraph IWA & $\mathbf{0.050}$ & $\mathbf{ 0.028 }$ \\
                & d & $11.7$ & $185.91\pm0.94$ & $-149.43\pm0.02$ & AO dark-hole, close & $\mathbf{0.050}$ & $\mathbf{ 0.023 }$  \\
\hline
               & b & $4.6$  & $137.17\pm0.97$ & $46.37\pm0.01$ & Low Wind Effet, close & $\mathbf{0.439}$ & $\mathbf{ 1.01 }$ \\
\bad{sphere2}  & c & $11.0$ & $218.54\pm0.80$ & $-41.84\pm0.42$ & Low Wind Effet, middle  & $\mathbf{0.049}$ & $\mathbf{ 0.542 }$ \\
               & d & $6.8$  & $291.85\pm0.79$ & $-116.63\pm0.09$ & Low Wind Effet, further & $\mathbf{0.480}$ & $\mathbf{ 0.411 }$ \\
\hline
                & b & $3.2$ & $161.59\pm2.08$ & $130.86\pm1.47$  & Coronagraph IWA & $\mathbf{0.074}$ & $\mathbf{ 0.812 }$ \\
$\multirow{-2}{*}{\nom{sphere3}}$  & c & $7.0$ & $456.35\pm1.01$ & $-79.78\pm0.18$  & AO ring of fire & $\mathbf{0.108}$ & $\mathbf{ 4.282 }$ \\
\hline
                & b & $6.5$ & $235.62\pm0.81$ & $-21.08\pm0.34$   & AO dark-hole close & $\mathbf{0.141}$ & $\mathbf{ 0.226 }$ \\
\bad{sphere4}   & c & $2.0$ & $430.10\pm0.32$ & $119.28\pm0.08$   & AO dark-hole far & $\mathbf{0.013}$ & $\mathbf{ > 500 }$ \\
                & d & $5.9$ & $162.39\pm2.03$ & $-129.55\pm0.10$  & Coronagraph IWA  & $\mathbf{0.266}$ & $\mathbf{ 8.549 }$ \\
\hline
\end{tabular}
\end{center}
\end{table} 

For both GPI and SPHERE-IFS, the observing conditions have an important impact on the quality and accuracy of the estimated astrometry. As expected, the baseline SNR of the injection also has an impact on both the estimation and their uncertainties (being larger with lower SNR). The detections close to the coronagraphic inner working angle (IWA) usually introduce a small bias in the estimation, except in the case of excellent observing conditions. When the low wind effect (LWE) is strong, as in the ~\emph{sphere2} data set, we observe a small bias in the estimated astrometry for the injection located on the diffraction patterns due to the LWE; with the exception of the one injection 'c' located in the middle of a diffraction pattern. 
Overall, the PCA-ASDI+NEGFC technique provides excellent results: the ground-truth falls within the 1-$\sigma$ uncertainties for about $50\%$ of the injections, and always within the 3-$\sigma$ uncertainties except for 2 injections (\emph{sphere2} data set affected by LWE). The astrometry estimated for the $21$ injected planetary signals includes the ground truth within the 5-$\sigma$ uncertainties. In addition, the estimated astrometry is always within one resolution element from the ground-truth.

\subsection{Astrometry: results as a function of the submitted algorithms}
The results for the baseline method and the four submitted methods are presented in Tab.~\ref{tab:resastrometry_gpi} (GPI datasets) and Tab.~\ref{tab:resastrometry_sphere} (SPHERE-IFS datasets). For each dataset and injection, we show the chosen metric\cite{eidc2022} as the euclidean distance between the estimation and the ground-truth. The last column averages the results over all injections to show an overall quantity for each method. A graphical view of the results is presented in Fig.~\ref{fig:astrom_gpi} and Fig.~\ref{fig:astrom_sphere}  for the GPI and SPHERE-IFS data respectively.

\begin{table}[!h]
\caption[example] 
   {\label{tab:resastrometry_gpi} 
Astrometry results on the four GPI data: euclidean distance between the estimated position and the ground-truth (both in Cartesian coordinates) for each injection within each data set. For the baseline method and the four submitted methods, the metric is shown for (i) each injection, (ii) averaged for each data set (bold values), and (iii) averaged on the four GPI data sets (green bold values).
}
\begin{center}
\begin{tabular}{|l|| c | c | c || c | c || c | c | c || c | c || c |} 
\hline
\multirow{2}{*}{\textbf{Method}} & \multicolumn{3}{c||}{\good{gpi1}} & \multicolumn{2}{c||}{\nom{gpi2}} & \multicolumn{3}{c||}{\nom{gpi3}} & \multicolumn{2}{c||}{\bad{gpi4}} & \multirow{2}{*}{\textbf{Mean}} \\ 
\cline{2-11}
& b & c & d & b & c  & b & c & d & b & c & \\
\hline\hline
$\multirow{2}{*}{$\mathrm{ Baseline }$}$
& $ 0.004 $
& $ 0.354 $
& $ 0.053 $
& $ 0.099 $
& $ 0.609 $
& $ 0.043 $
& $ 0.138 $
& $ 0.004 $
& $ 0.024 $
& $ 0.134 $
& \\ \cline{2-11}
& \multicolumn{3}{c||}{$ \mathbf{ 0.137 }$}
& \multicolumn{2}{c||}{$ \mathbf{ 0.354 }$}
& \multicolumn{3}{c||}{$ \mathbf{ 0.062 }$}
& \multicolumn{2}{c||}{$ \mathbf{ 0.079 }$}
& \multirow{-2}{*}{\good{$ \mathbf{ 0.15 }$}}
\\
\hline
$\multirow{2}{*}{$\mathrm{ RSM }$}$
& $ 0.453 $
& $ 0.779 $
& $ 0.053 $
& $ 0.127 $
& $ 0.534 $
& $ 0.049 $
& $ 0.727 $
& $ 0.119 $
& $ 0.126 $
& $ 0.324 $
& \\ \cline{2-11}
& \multicolumn{3}{c||}{$ \mathbf{ 0.428 }$}
& \multicolumn{2}{c||}{$ \mathbf{ 0.33 }$}
& \multicolumn{3}{c||}{$ \mathbf{ 0.298 }$}
& \multicolumn{2}{c||}{$ \mathbf{ 0.225 }$}
& \multirow{-2}{*}{\good{$ \mathbf{ 0.33 }$}}
\\
\hline
$\multirow{2}{*}{$\mathrm{ pyKLIP }$}$
& $ 0.088 $
& $ 0.1 $
& $ 0.035 $
& $ 0.018 $
& $ 0.302 $
& $ 0.137 $
& $ 0.325 $
& $ 0.169 $
& $ 0.176 $
& $ 1.05 $
& \\ \cline{2-11}
& \multicolumn{3}{c||}{$ \mathbf{ 0.074 }$}
& \multicolumn{2}{c||}{$ \mathbf{ 0.16 }$}
& \multicolumn{3}{c||}{$ \mathbf{ 0.21 }$}
& \multicolumn{2}{c||}{$ \mathbf{ 0.613 }$}
& \multirow{-2}{*}{\good{$ \mathbf{ 0.24 }$}}
\\
\hline
$\multirow{2}{*}{$\mathrm{ AMAT }$}$
& $ 0.16 $
& $ 0.268 $
& $ 0.079 $
& $ 0.102 $
& $ 0.69 $
& $ 0.275 $
& $ 0.175 $
& $ 0.03 $
& $ 0.075 $
& $ 0.131 $
& \\ \cline{2-11}
& \multicolumn{3}{c||}{$ \mathbf{ 0.169 }$}
& \multicolumn{2}{c||}{$ \mathbf{ 0.396 }$}
& \multicolumn{3}{c||}{$ \mathbf{ 0.16 }$}
& \multicolumn{2}{c||}{$ \mathbf{ 0.103 }$}
& \multirow{-2}{*}{\good{$ \mathbf{ 0.20 }$}}
\\
\hline
$\multirow{2}{*}{$\mathrm{ ANDRO }$}$
& $ 3.928 $
& $ 2.563 $
& $ 4.387 $
& $ 5.28 $
& $ 4.393 $
& $ 8.827 $
& $ 6.569 $
& $ 4.617 $
& $ 1.389 $
& $ 5.819 $
& \\ \cline{2-11}
& \multicolumn{3}{c||}{$ \mathbf{ 3.626 }$}
& \multicolumn{2}{c||}{$ \mathbf{ 4.837 }$}
& \multicolumn{3}{c||}{$ \mathbf{ 6.671 }$}
& \multicolumn{2}{c||}{$ \mathbf{ 3.604 }$}
& \multirow{-2}{*}{\good{$ \mathbf{ 4.78 }$}}
\\
\hline
\end{tabular}
\end{center}
\end{table}

\begin{table}[!h]
\caption[example] 
   {\label{tab:resastrometry_sphere} 
Astrometry results on the four SPHERE-IFS data: euclidean distance between the estimated position and the ground-truth (both in Cartesian coordinates) for each injection within each data set. For the baseline method and the four submitted methods, the metric is shown for (i) each injection, (ii) averaged for each data set (bold values), and (iii) averaged on the four SPHERE-IFS data sets (green bold values).
}
\begin{center}
\begin{tabular}{|l|| c | c | c || c | c | c || c | c || c | c | c || c |} 
\hline
\multirow{2}{*}{\textbf{Method}} & \multicolumn{3}{c||}{\good{sphere1}} & \multicolumn{3}{c||}{\bad{sphere2}} & \multicolumn{2}{c||}{\nom{sphere3}} & \multicolumn{3}{c||}{\bad{sphere4}} & \multirow{2}{*}{\textbf{Mean}} \\ 
\cline{2-12}
& b & c & d & b & c & d  & b & c  & b & c & d & \\
\hline\hline
$\multirow{2}{*}{$\mathrm{ Baseline }$}$
& $ 0.035 $
& $ 0.05 $
& $ 0.05 $
& $ 0.44 $
& $ 0.049 $
& $ 0.48 $
& $ 0.074 $
& $ 0.108 $
& $ 0.141 $
& $ 0.013 $
& $ 0.266 $
& \\ \cline{2-12}
& \multicolumn{3}{c||}{$ \mathbf{ 0.045 }$}
& \multicolumn{3}{c||}{$ \mathbf{ 0.323 }$}
& \multicolumn{2}{c||}{$ \mathbf{ 0.091 }$}
& \multicolumn{3}{c||}{$ \mathbf{ 0.14 }$}
& \multirow{-2}{*}{\good{$ \mathbf{ 0.16 }$}}
\\
\hline
$\multirow{2}{*}{$\mathrm{ RSM }$}$
& $ 0.343 $
& $ 0.191 $
& $ 0.128 $
& $ 0.828 $
& $ 0.297 $
& $ 0.297 $
& $ 0.254 $
& $ 1.835 $
& $ 0.185 $
& $ 0.036 $
& $ 0.638 $
& \\ \cline{2-12}
& \multicolumn{3}{c||}{$ \mathbf{ 0.221 }$}
& \multicolumn{3}{c||}{$ \mathbf{ 0.474 }$}
& \multicolumn{2}{c||}{$ \mathbf{ 1.045 }$}
& \multicolumn{3}{c||}{$ \mathbf{ 0.286 }$}
& \multirow{-2}{*}{\good{$ \mathbf{ 0.46 }$}}
\\
\hline
$\multirow{2}{*}{$\mathrm{ pyKLIP }$}$
& $ 0.184 $
& $ 0.071 $
& $ 0.293 $
& $ 0.38 $
& $ 0.23 $
& $ 0.554 $
& $ 0.231 $
& $ 0.054 $
& $ 0.092 $
& $ 0.068 $
& $ 0.052 $
& \\ \cline{2-12}
& \multicolumn{3}{c||}{$ \mathbf{ 0.183 }$}
& \multicolumn{3}{c||}{$ \mathbf{ 0.388 }$}
& \multicolumn{2}{c||}{$ \mathbf{ 0.142 }$}
& \multicolumn{3}{c||}{$ \mathbf{ 0.071 }$}
& \multirow{-2}{*}{\good{$ \mathbf{ 0.20 }$}}
\\
\hline
$\multirow{2}{*}{$\mathrm{ AMAT }$}$
& $ 0.223 $
& $ 0.58 $
& $ 0.037 $
& $ 0.506 $
& $ 0.226 $
& $ 0.325 $
& $ 0.362 $
& $ 0.01 $
& $ 0.085 $
& $ 0.109 $
& $ 0.238 $
& \\ \cline{2-12}
& \multicolumn{3}{c||}{$ \mathbf{ 0.28 }$}
& \multicolumn{3}{c||}{$ \mathbf{ 0.353 }$}
& \multicolumn{2}{c||}{$ \mathbf{ 0.186 }$}
& \multicolumn{3}{c||}{$ \mathbf{ 0.144 }$}
& \multirow{-2}{*}{\good{$ \mathbf{ 0.25 }$}}
\\
\hline
$\multirow{2}{*}{$\mathrm{ ANDRO }$}$
& $ 2.292 $
& $ 3.236 $
& $ 2.746 $
& $ 1.102 $
& $ 0.871 $
& $ 3.394 $
& $ 5.103 $
& $ 2.07 $
& $ 2.702 $
& $ 5.389 $
& $ 6.166 $
& \\ \cline{2-12}
& \multicolumn{3}{c||}{$ \mathbf{ 2.758 }$}
& \multicolumn{3}{c||}{$ \mathbf{ 1.789 }$}
& \multicolumn{2}{c||}{$ \mathbf{ 3.587 }$}
& \multicolumn{3}{c||}{$ \mathbf{ 4.752 }$}
& \multirow{-2}{*}{\good{$ \mathbf{ 3.19 }$}}
\\
\hline
\end{tabular}
\end{center}
\end{table} 

The presented numbers and appendix figures lead to the following conclusions:\\
-Overall, higher SNR lead to a better estimation of the position, as expected;\\
-Apart of ANDROMEDA, the estimations are rarely off by more than one resolution element; \\
-Observing conditions do not seem to be the major factor in the accuracy of the position estimation. Instead the localisation seem to be the leading factor. Indeed, the proximity to any bright feature due to a diffraction effect (such as the AO-correction ring, satellite spots, apodized-Lyot coronagraph inner-working angle, low wind effect or so) varying with the wavelength might affect the overall estimated position of the injection by distorting its flux distribution depending on the spectral channel.\\
-On the contrary, even at low SNR, but at a favorable location (in the AO-corrected area, at large separation from the star), the estimations are very accurate (e.g. \emph{sphere4}, injection 'c').\\
-We note here the specific case of strong low wind effect in the \emph{sphere2} data set. The inner ('b') and outer ('d') injections, both sitting on an intense spider diffraction pattern, are not as well estimated as the other injection ('c') although located right on a spider diffraction pattern. The better estimation for the latter may also be accounted for a higher SNR.

In terms of methods, most trends are similar from one injection to another and from one data set to another (at a few exceptions). Overall, PCA-SADI+NEGFC, AMAT, RSM and pyKLIP-FM all succeed in retrieving the injection position with a very good accuracy, usually within one resolution element. Only the classical version of ANDROMEDA, fitting a 2-D Gaussian at the location of a detection in the SNR map, shows inaccurate astrometry estimation often exceeding a few pixels.

\section{Results: Spectrophotometry}
As a reminder, for the spectro-photometry estimation of the Exoplanet Imaging Data Challenge second phase, we asked the participants to provide with (1) the estimated contrast of each injected signal at each wavelength, (2) optionally the corresponding 1$\sigma$ uncertainties of the estimations, and (3) optionally the corresponding posterior distribution used to estimate the contrast and its uncertainties. As for the metric used for the ranking of the submitted results, the EIDC phase 2 working group decided to compute the normalized absolute distance (in the sense of the L1-norm) between the estimated value and the ground truth value for each spectral channel separately. Normalizing the absolute distance by the ground-truth contrast ensures that low-contrast signals do not penalize smaller true contrasts in the averaged score. 
More details about the data, submission process and metrics for comparison and ranking can be found in our previous SPIE publication Cantalloube et al., 2022\cite{cantalloube2022eidc}.

\subsection{Spectro-photometry: baseline results as a function of the dataset}
As a first qualitative interpretation, we compare the spectro-photometry estimated by the baseline PCA-SADI+NEGFC algorithm in each data set. The results on the GPI images are shown in Fig.~\ref{fig:data_spec_gpi} and those on the SPHERE-IFS images are shown in Fig.~\ref{fig:data_spec_sph}. On the figure we show on the left the actual position of the injections overlaid in the first coronagraphic image of the cube and on the right the extracted spectra for each injection (2 to 3 injection per dataset) including the 3$\sigma$ errorbars compared to the ground truth (dark solid line), as well as the residuals between the extracted spectra and the corresponding ground-truth (shaded areas represent the corresponding 3$\sigma$ uncertainties). 
Quantitative results, including the comparison metric used, are presented in Tab.~\ref{tab:agpi} and Tab.~\ref{tab:asph} for the GPI and SPHERE-IFS data respectively.

Overall, there are a few singular outliers at some random wavelength of the estimated spectrum (probably due to some convergence problems in the NEGFC method). Seemingly, there is no systematic wavelength affected by this problem. Regardless of these outliers, the spectro-photometric estimations always include the ground-truth within 5$\sigma$. Uncertainties are matching the spectrum depth: lower contrast yield larger uncertainties, as expected (clearly visible in e.g. companion 'c' of data \emph{gpi1}, \emph{gpi3} and \emph{sphere1}). 

Globally, the further the injected companion, the better its spectro-photometric estimation. Higher SNR yield better estimations. Equivalently, injections at lower contrast are better estimated (see e.g. \emph{sphere1} 'b' and \emph{gpi1} 'c', both having a very high contrast, under exquisite observing conditions and located right in the AO-corrected zone). In general, the structure of the spectrum is correctly retrieved. 

There are two notable exceptions to these conclusions, due to poor observation conditions. Low-wind effect (LWE) is present in the \emph{sphere2} data set, in which we placed three injections on top of the characteristics spider diffraction peaks. As a result, the three corresponding spectra are underestimated (lower than the ground-truth): the presence of low-wind effect biases the spectrum estimation. Wind-driven halo (WDH) is present in the \emph{gpi2} and the \emph{sphere4} data set and we placed injections \emph{gpi2} 'b' and \emph{sphere4} 'd' within the characteristic butterfly shape. The estimated spectrum of \emph{gpi2} 'b' (located beyond $600\mathrm{mas}$), does not seem to be much affected, but \emph{sphere4} 'd' estimated spectrum is overestimated despite a relatively good SNR: the presence of wind-driven halo seems to affect close-in companions, but this needs to be verified in a more systematic fashion to be confirmed.

\subsection{Spectro-photometry: results as a function of the submitted algorithms}
The results for the baseline method and the three submitted methods are presented in Tab.~\ref{tab:resphotometry_gpi} (GPI datasets) and Tab.~\ref{tab:resphotometry_sph} (SPHERE-IFS datasets). For each dataset and injection, we show the chosen metric\cite{eidc2022} as the normalized L1-norm distance between the estimation to the ground-truth. The last column averages the results for each injection to present an overall quantity for each algorithm. A graphical view of the results is presented in Fig.~\ref{fig:photom_gpi} and Fig.~\ref{fig:photom_sphere} for the GPI and SPHERE-IFS data respectively.

On average over all the injections, at all wavelength and for all the data set, it appears that PCA-NEGFC and pyKLIP-FM are better at retrieving the spectro-photometry, followed closely by AMAT, then RSM. These trends remain the same with various observing conditions. For the GPI data set, all methods give metrics below one but RSM and PCA-NEGFC for the closest companion of the \emph{gpi4} data set (under bad observing conditions) showing a SNR below 5. For the SPHERE-IFS data set, the very high values obtained for \emph{sphere4} 'c' are due to a very low SNR of the injection (less than 2 for the baseline). 

However, when we look at the detailed extractions shown on Fig.~\ref{fig:photom_gpi} and Fig.~\ref{fig:photom_sphere} (for GPI and SPHERE-IFS respectively), we may notice a few additional effects hidden in the averaged metric. 
Some algorithms are good for faint planets (e.g. pyKLIP and AMAT), but suffer from significant biases when it comes to relatively bright planets, and reversely the baseline can yield significant outliers for faint planets but is very reliable for bright planets. 
The adopted procedure for the \textbf{PCA-NEGFC} (baseline), which selects the number of principal components that optimize the SNR may be prone to overestimating the flux in some channels, by including a residual speckle signal. 
\textbf{pyKLIP-FM} seems to overestimate the flux at lower contrast values (bright planets): the oversubtraction may not be well estimated for channels where the companion is bright. In general the structure of the extracted spectra appear much smoother compared to e.~g. the baseline. 
\textbf{AMAT} is relatively good in general, but slightly underestimates the flux, in particular under good observation conditions and/or for bright planets. 
\textbf{RSM} is still good on average but suffers from many outliers and the spectrum structure is jagged. 
Note that the results concerning pyKLIP-FM and VIP/PCA-NEGFC are similar to the conclusions presented in Nasedkin et al.\cite{nasedkin2023pp} (2023) for pyKLIP-FM vs Pynpoint/PCA-NEGFC.

In the specific case of low-wind effect (\emph{sphere2} data set), all algorithm are underestimating the spectra mainly at larger wavelengths. Only RSM (green lines) and AMAT (blue lines) are estimating correctly at short wavelengths for only one or two in three injections.

\begin{table}[!h]
\caption[example] 
   {\label{tab:resphotometry_gpi} 
Spectro-photometry results on the four GPI data set: absolute distance between the estimated photometry at each spectral channel and the corresponding ground-truth for each injection within each data set, normalized by the ground-truth. For the baseline method and the four submitted methods, the metric is shown for (i) each injection, (ii) averaged over each injection (bold values), and (iii) averaged on the four GPI data sets (green bold values).
}
\begin{center}
\begin{tabular}{|l|| c | c | c || c | c || c | c | c || c | c || c |} 
\hline
\multirow{2}{*}{\textbf{Method}} & \multicolumn{3}{c||}{\good{gpi1}}& \multicolumn{2}{c||}{\nom{gpi2}} & \multicolumn{3}{c||}{\nom{gpi3}} & \multicolumn{2}{c||}{\bad{gpi4}} & \multirow{2}{*}{\textbf{Median}} \\ 
\cline{2-11}
& b & c & d & b & c  & b & c & d & b & c & \\
\hline\hline
$\multirow{2}{*}{$\mathrm{Baseline}$}$ 
& 0.559
& 0.423
& 0.023
& 0.067
& 0.209
& 0.073
& 0.18
& 0.013
& 0.212
& 3.003
& \\ \cline{2-11}
& \multicolumn{3}{c||}{$ \mathbf{ 0.335 }$}
& \multicolumn{2}{c||}{$ \mathbf{ 0.138 }$}
& \multicolumn{3}{c||}{$ \mathbf{ 0.089 }$}
& \multicolumn{2}{c||}{$ \mathbf{ 1.607 }$}
& \multirow{-2}{*}{\good{$ \mathbf{ 0.237 }$}}
\\
\hline
$\multirow{2}{*}{$\mathrm{RSM}$}$ 
& 0.822
& 0.307
& 0.077
& 0.116
& 0.474
& 0.091
& 0.213
& 0.068
& 0.094
& 3.456
& \\ \cline{2-11}
& \multicolumn{3}{c||}{$ \mathbf{ 0.402 }$}
& \multicolumn{2}{c||}{$ \mathbf{ 0.295 }$}
& \multicolumn{3}{c||}{$ \mathbf{ 0.124 }$}
& \multicolumn{2}{c||}{$ \mathbf{ 1.775 }$}
& \multirow{-2}{*}{\good{$ \mathbf{ 0.349 }$}}
\\
\hline
$\multirow{2}{*}{$\mathrm{pyKLIP}$}$ 
& 0.251
& 0.033
& 0.226
& 0.449
& 0.204
& 0.34
& 0.183
& 0.308
& 0.25
& 0.204
& \\ \cline{2-11}
& \multicolumn{3}{c||}{$ \mathbf{ 0.17 }$}
& \multicolumn{2}{c||}{$ \mathbf{ 0.327 }$}
& \multicolumn{3}{c||}{$ \mathbf{ 0.277 }$}
& \multicolumn{2}{c||}{$ \mathbf{ 0.227 }$}
& \multirow{-2}{*}{\good{$ \mathbf{ 0.252 }$}}
\\
\hline
$\multirow{2}{*}{$\mathrm{AMAT}$}$ 
& 0.663
& 0.11
& 0.213
& 0.165
& 0.46
& 0.085
& 0.276
& 0.026
& 0.122
& 0.105
& \\ \cline{2-11}
& \multicolumn{3}{c||}{$ \mathbf{ 0.329 }$}
& \multicolumn{2}{c||}{$ \mathbf{ 0.312 }$}
& \multicolumn{3}{c||}{$ \mathbf{ 0.129 }$}
& \multicolumn{2}{c||}{$ \mathbf{ 0.114 }$}
& \multirow{-2}{*}{\good{$ \mathbf{ 0.221 }$}}
\\
\hline
\end{tabular}
\end{center}
\end{table}

\begin{table}[!h]
\caption[example] 
   {\label{tab:resphotometry_sph} 
Spectro-photometry results on the four SPHERE-IFS data: absolute distance between the estimated photometry at each spectral channel and the corresponding ground-truth for each injection within each data set, normalized by the ground-truth. For the baseline method and the four submitted methods, the metric is shown for (i) each injection, (ii) averaged over each injection (bold values), and (iii) averaged on the four GPI data sets (green bold values). 
}
\vspace{-0.4cm}
\begin{center}
\begin{tabular}{|l|| c | c | c || c | c | c || c | c || c | c | c || c |} 
\hline
\multirow{2}{*}{\textbf{Method}} & \multicolumn{3}{c||}{\good{sphere1}} & \multicolumn{3}{c||}{\bad{sphere2}} & \multicolumn{2}{c||}{\nom{sphere3}} & \multicolumn{3}{c||}{\bad{sphere4}} & \multirow{2}{*}{\textbf{Median}} \\ 
\cline{2-12}
& b & c & d & b & c & d  & b & c  & b & c & d & \\
\hline\hline
$\multirow{2}{*}{$\mathrm{Baseline}$}$ 
& 1.618
& 0.028
& 0.023
& 1.01
& 0.542
& 0.411
& 0.812
& 4.28
& 0.226
& $>500$
& 8.55
& \\ \cline{2-12}
& \multicolumn{3}{c||}{$ \mathbf{ 0.556 }$}
& \multicolumn{3}{c||}{$ \mathbf{ 0.654 }$}
& \multicolumn{2}{c||}{$ \mathbf{ 2.547 }$}
& \multicolumn{3}{c||}{$ \mathbf{ > 1000 }$}
& \multirow{-2}{*}{\good{$ \mathbf{ 1.60 }$}}
\\
\hline
$\multirow{2}{*}{$\mathrm{RSM}$}$ 
& 1.78
& 0.079
& 0.053
& 6.32
& 0.104
& 0.3
& 0.361
& 14.5
& 0.272
& $>500$
& 0.408
& \\ \cline{2-12}
& \multicolumn{3}{c||}{$ \mathbf{ 0.64 }$}
& \multicolumn{3}{c||}{$ \mathbf{ 2.24 }$}
& \multicolumn{2}{c||}{$ \mathbf{ 7.43 }$}
& \multicolumn{3}{c||}{$ \mathbf{ 355.53 }$}
& \multirow{-2}{*}{\good{$ \mathbf{ 4.83 }$}}
\\
\hline
$\multirow{2}{*}{$\mathrm{pyKLIP}$}$ 
& 0.199
& 0.239
& 0.099
& 0.572
& 0.599
& 0.524
& 0.356
& 6.08
& 0.044
&$>500$
& 0.114
& \\ \cline{2-12}
& \multicolumn{3}{c||}{$ \mathbf{ 0.179 }$}
& \multicolumn{3}{c||}{$ \mathbf{ 0.565 }$}
& \multicolumn{2}{c||}{$ \mathbf{ 3.22 }$}
& \multicolumn{3}{c||}{$ \mathbf{ 207.64 }$}
& \multirow{-2}{*}{\good{$ \mathbf{ 1.89 }$}}
\\
\hline
$\multirow{2}{*}{$\mathrm{AMAT}$}$ 
& 1.48
& 0.029
& 0.118
& 0.175
& 0.242
& 0.294
& 0.273
& 8.57
& 0.139
& $>500$
& 0.123
& \\ \cline{2-12}
& \multicolumn{3}{c||}{$ \mathbf{ 0.541 }$}
& \multicolumn{3}{c||}{$ \mathbf{ 0.237 }$}
& \multicolumn{2}{c||}{$ \mathbf{ 4.42 }$}
& \multicolumn{3}{c||}{$ \mathbf{ 235.84 }$}
& \multirow{-2}{*}{\good{$ \mathbf{ 2.48 }$}}
\\
\hline
\end{tabular}
\end{center}
\end{table}

\section{CONCLUSION \& LEGACY}
The \emph{Exoplanet Imaging Data Challenge} is a community-led project that aims at offering tools for a better understanding of the current capacities of the various high-contrast imaging post-processing techniques, and to foster developments of new methods. To that end, we offer several resources and an extensive bibliography gathered on our website. As part of the data challenge, we provide with a curated set of data that are accessible on a \emph{Zenodo} repository, as well as a set of assorted metrics to gauge the results of a given algorithm that are available on a \emph{Github} repository. In addition, to ease the comparison to other methods, we wrote a \emph{jupyter notebook} that processes any given input and provides the comparison graphics and ranking with respect to the results already published in the 1st phase of the data challenge (Cantalloube et al.\cite{Cantalloube2020eidc}, 2020). 

Since its creation in 2019, several publications have used the data set and metric set to support the results of new post-processing methods. After several years of running this data challenge, we have gathered some \emph{lessons learned} to improve the engagement of participants. This lack of engagement turns out to be the major breakpoint preventing us from collecting a sufficient number of inputs in order to compile a consistent comparative conclusion to communicate the community. One important piece of advice is to organize a special hands-on session to help with the basics of the data format. Another idea we tried is to involve students during lab sessions. We also reached out to networks that run data challenges in other scientific fields to get feedback and ensure a smooth management of the EIDC.

To enhance this legacy prospect, in the future we would like to gather more data from the state-of-the-art high-contrast imaging instruments, such as Subaru/CHARIS, Magellan-Clay/MagAO-X, VLT/NEAR, LBT/SHARK-NIR and LBT/SHARK-VIS, under various observing conditions. These data would come with a set of tools to inject planetary signals and circumstellar disks, as well as metrics for comparison. As a community project based on volunteer work, we are doing our best to facilitate the use of the EIDC resources. For any suggestion or specific request, please contact us at \url{exoimg.datachallenge@gmail.com}. 

\appendix    

\section{Gallery: Baseline results}
\label{sec:gallery_bsl}

\begin{figure}
    \centering
    \resizebox{\hsize}{!}{\includegraphics{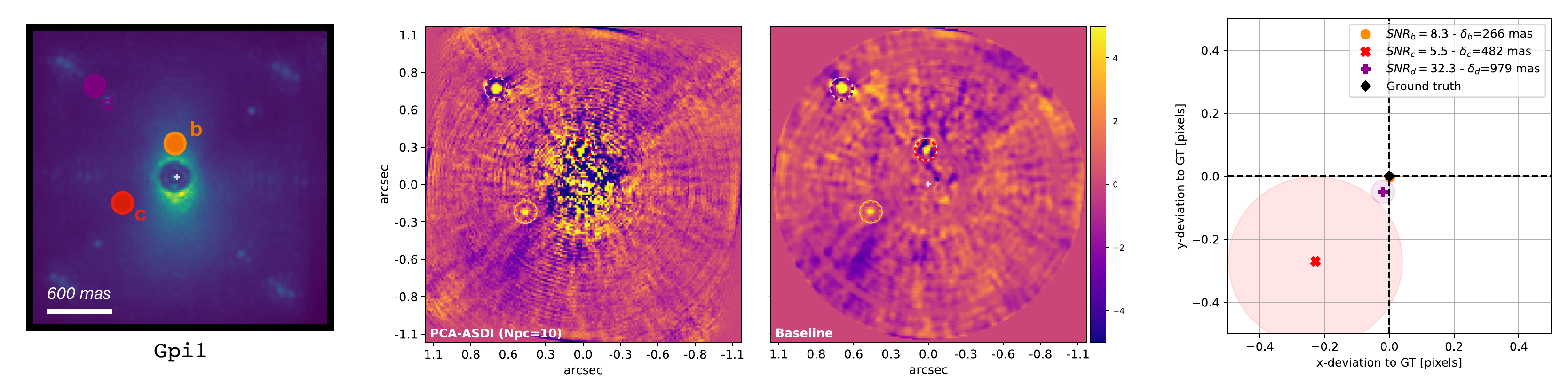}}
    \resizebox{\hsize}{!}{\includegraphics{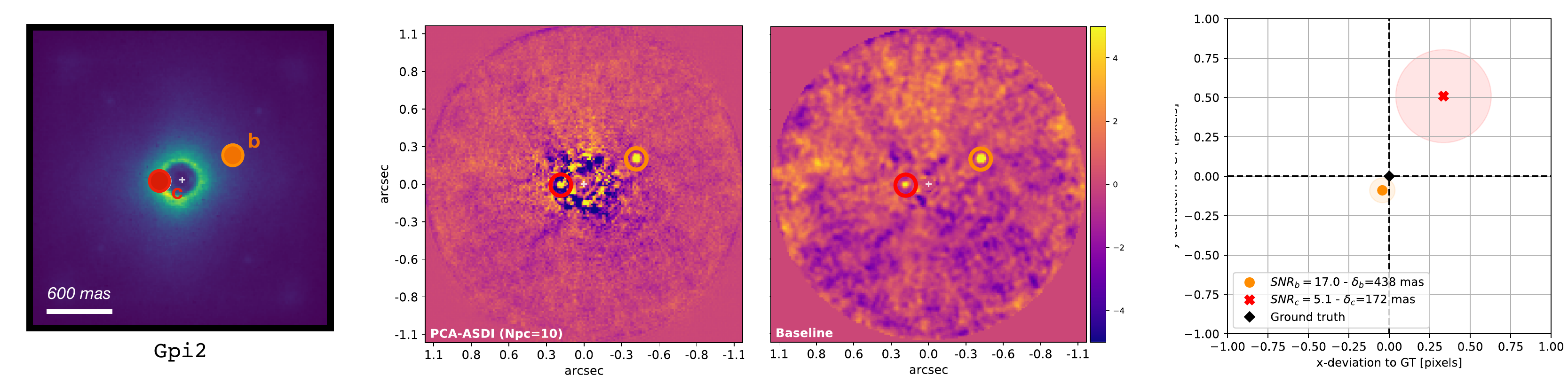}}
    \resizebox{\hsize}{!}{\includegraphics{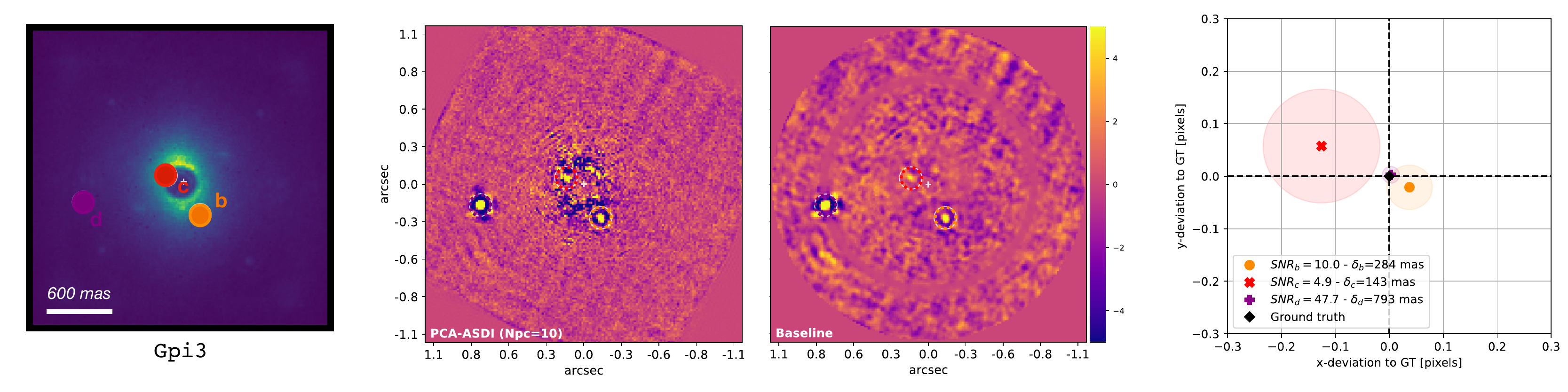}}
    \resizebox{\hsize}{!}{\includegraphics{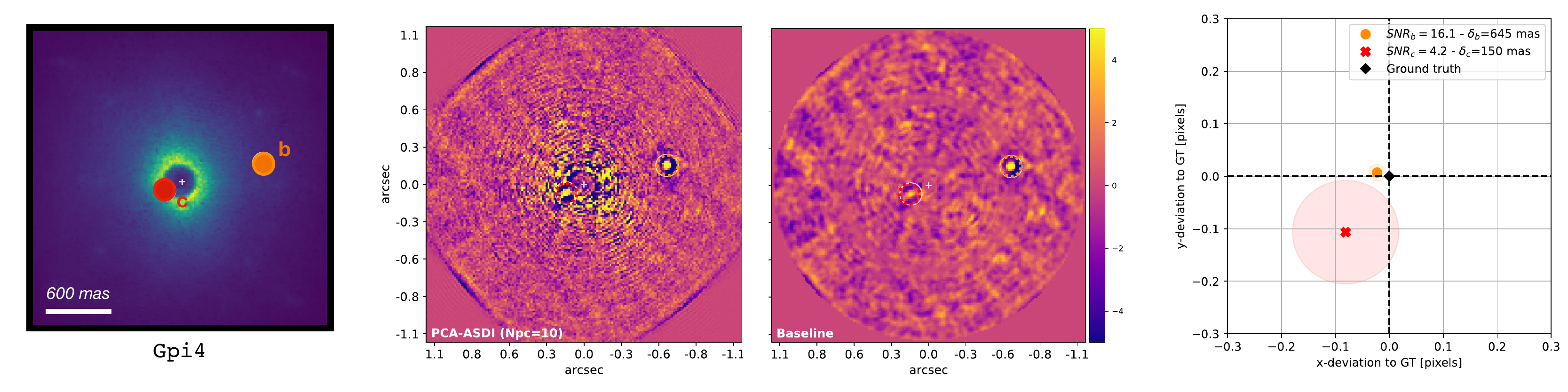}}
    \vspace{0.2cm} 
    \caption{Astrometry: GPI data baseline results for each injection. On each image, the location of the injected synthetic planetary signals are shown with the orange, red and purple circles. From left to right: Temporal  median image at the shortest wavelength (left); residual map after applying a PCA-SADI post-processing, as implemented in the VIP package and using the first 10 principal components to build the reference image for subtraction (middle-left); corresponding SNR map, as implemented in the VIP package (middle-right); and astrometry estimations for each injections, using the baseline NEGFC technique (right). The latter highlights the relative astrometry errors (in pixels) for each injection (color coded) compared to the ground-truth shown as a black diamond in the center of the image. The shaded area correspond to the 1-sigma uncertainty on the astrometry estimation. From top to bottom: data set \textit{gpi1}, \textit{gpi2}, \textit{gpi3}, and \textit{gpi4}.}
    \label{fig:data_gpi}
\end{figure}

\begin{figure}
    \centering
    \resizebox{\hsize}{!}{\includegraphics{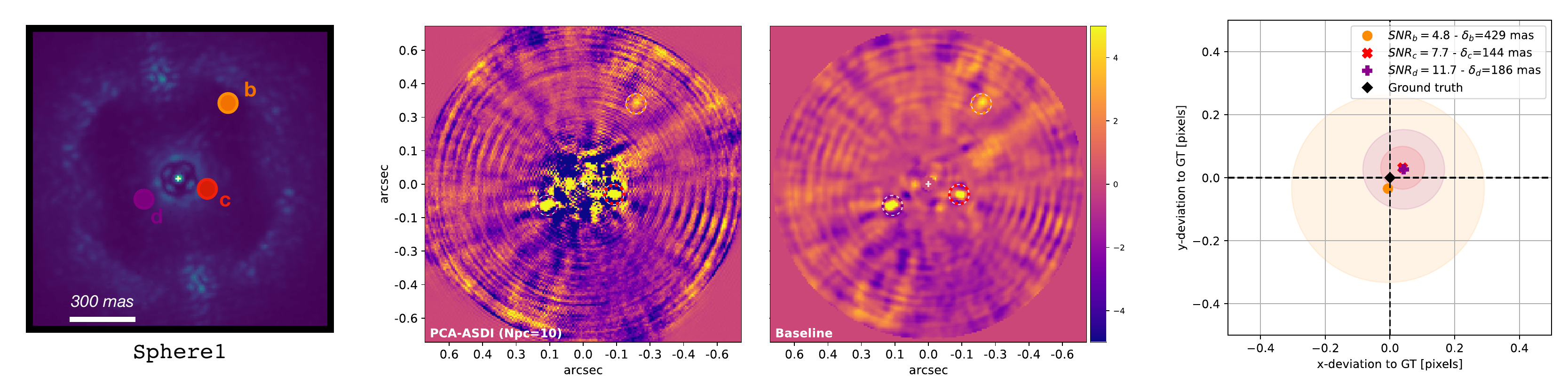}}
    \resizebox{\hsize}{!}{\includegraphics{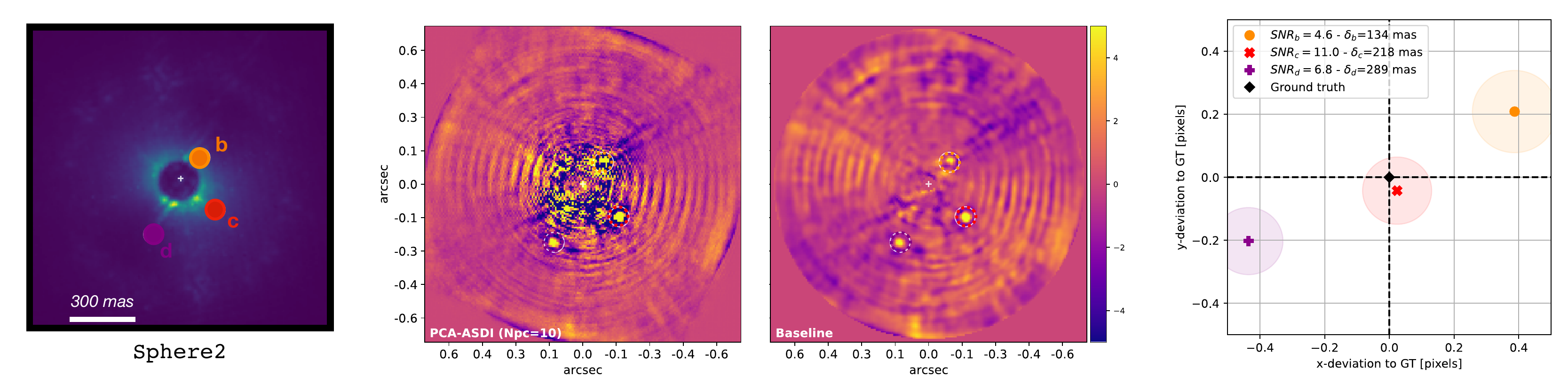}}
    \resizebox{\hsize}{!}{\includegraphics{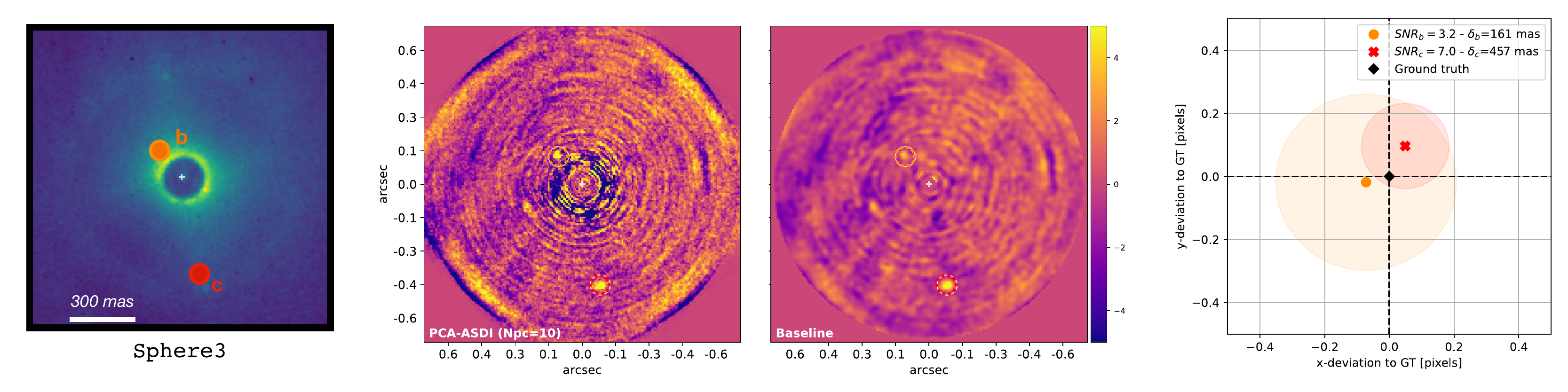}}
    \resizebox{\hsize}{!}{\includegraphics{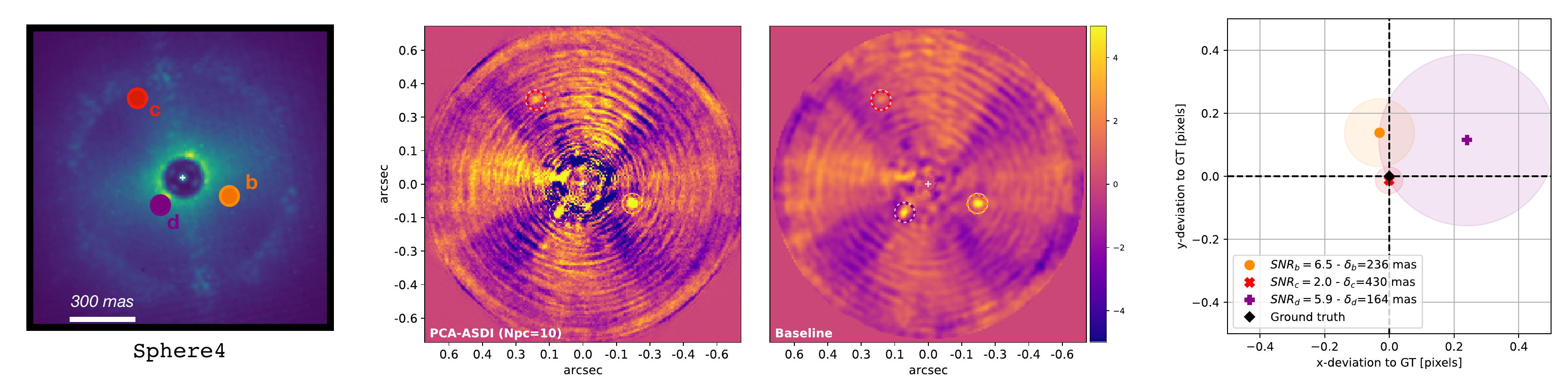}}
    \vspace{0.2cm} 
    \caption{Astrometry: SPHERE-IFS data baseline results for each injection. On each image, the location of the injected synthetic planetary signals are shown with the orange, red and purple circles. From left to right: Temporal  median image at the shortest wavelength (left); residual map after applying a PCA-SADI post-processing, as implemented in the VIP package and using the first 10 principal components to build the reference image for subtraction (middle-left); corresponding SNR map, as implemented in the VIP package (middle-right); and astrometry estimations for each injections, using the baseline NEGFC technique (right). The latter highlights the relative astrometry errors (in pixels) for each injection (color coded) compared to the ground-truth shown as a black diamond in the center of the image. The shaded area correspond to the 1-sigma uncertainty on the astrometry estimation. 
    From top to bottom: data set \textit{sphere1}, \textit{sphere2}, \textit{sphere3}, and \textit{sphere4}.}
    \label{fig:data_sphere}
\end{figure}

\begin{figure}
    \centering
    \resizebox{\hsize}{!}{\includegraphics{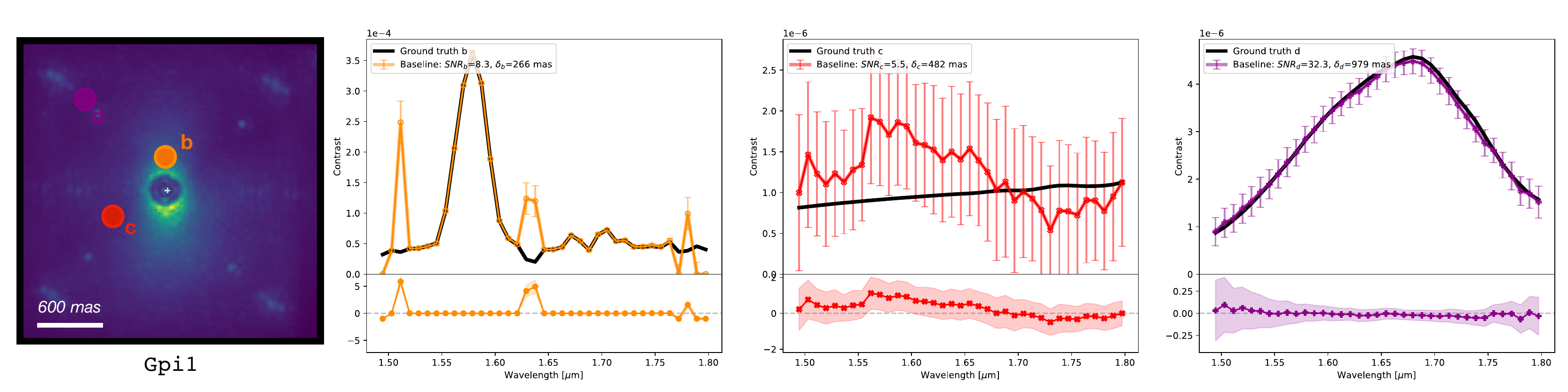}}
    \resizebox{\hsize}{!}{\includegraphics{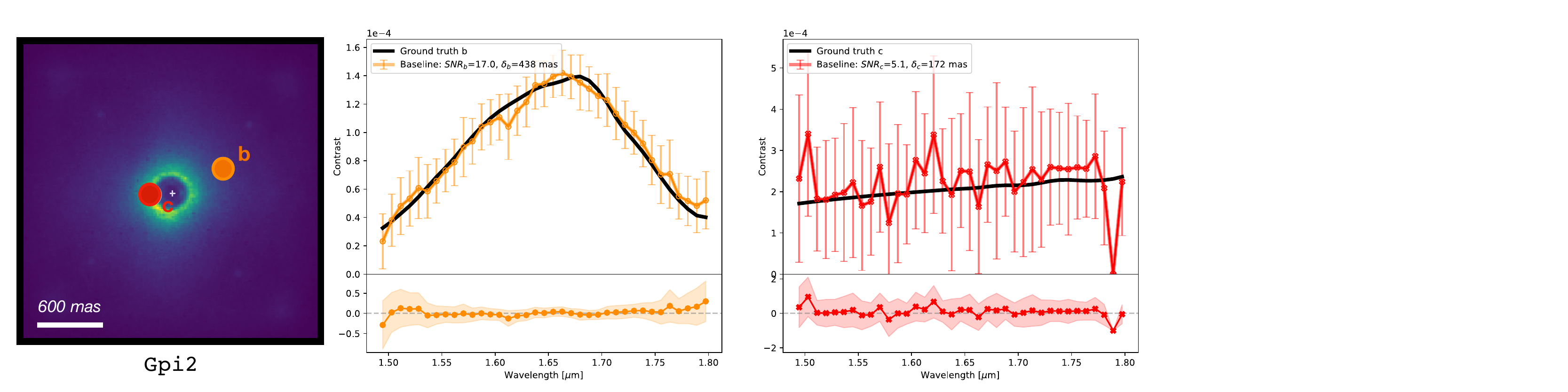}}
    \resizebox{\hsize}{!}{\includegraphics{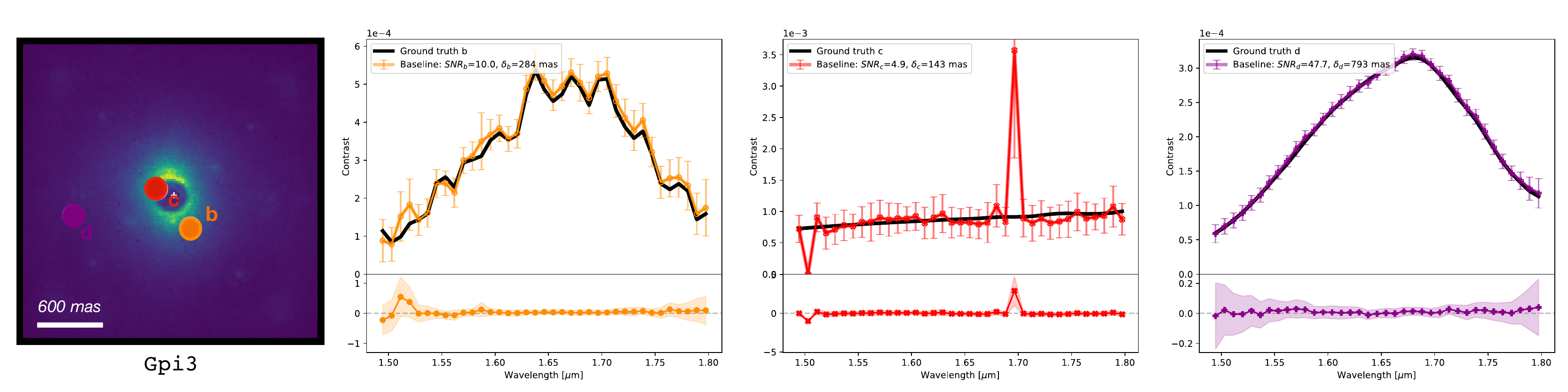}}
    \resizebox{\hsize}{!}{\includegraphics{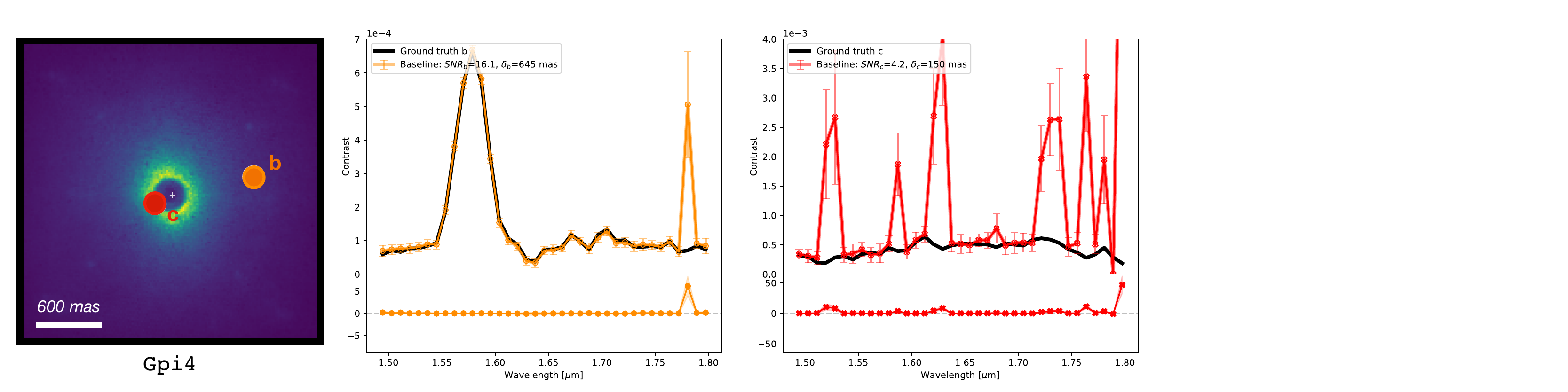}}
    \vspace{0.2cm} 
    \caption{Spectro-photometry: baseline results for each injection in the GPI data sets. From left to right: 
    Temporal  median image at the shortest wavelength with the position of the injected planetary signals (colored circles); Corresponding estimated spectrum with 3$\sigma$ uncertainties (colored lines), compared to the injected spectrum (dark line). The bottom panel shows the residuals between the estimated spectra and the ground-truth (the shaded area corresponds to the 3$\sigma$ uncertainties on the estimations).
    From top to bottom: data set \textit{gpi1}, \textit{gpi2}, \textit{gpi3}, and \textit{gpi4}.}
    \label{fig:data_spec_gpi}
\end{figure}

\begin{figure}
    \centering
    \resizebox{\hsize}{!}{\includegraphics{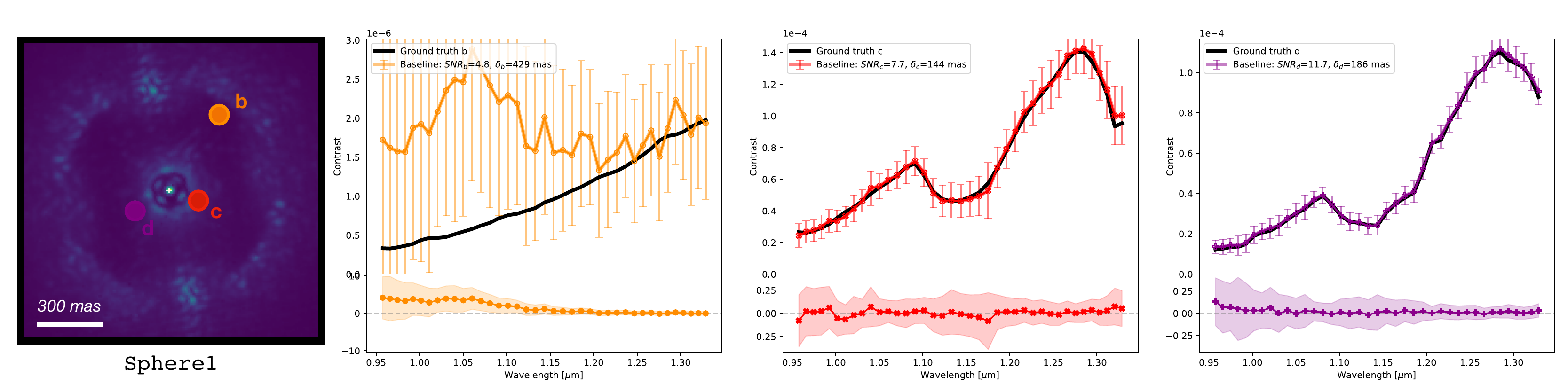}}
    \resizebox{\hsize}{!}{\includegraphics{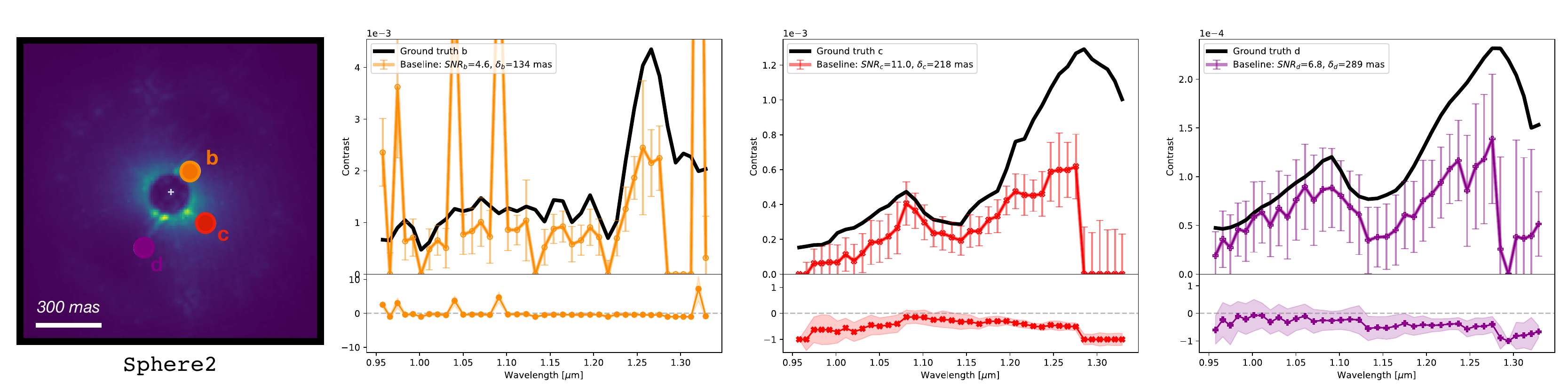}}
    \resizebox{\hsize}{!}{\includegraphics{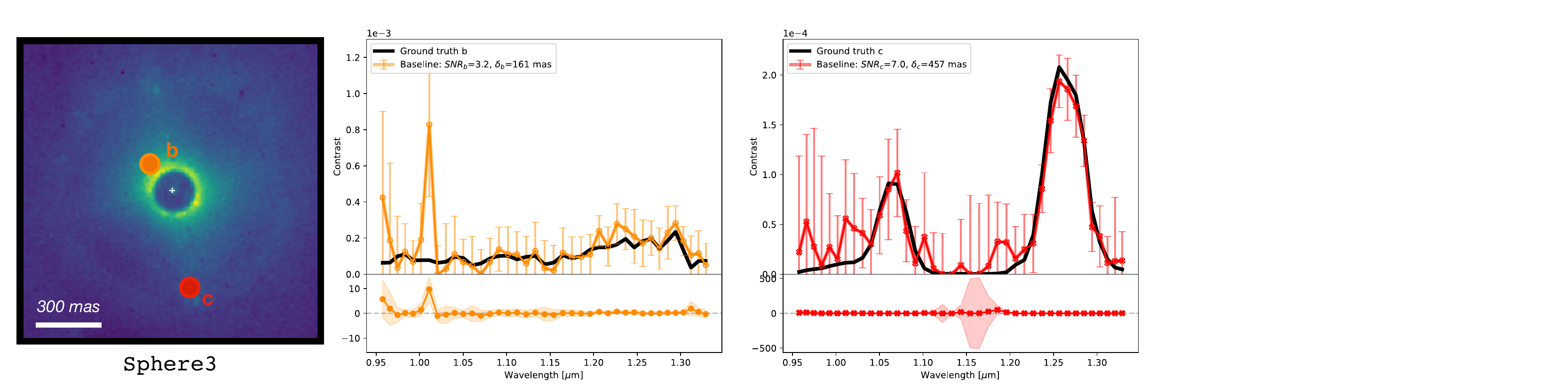}}
    \resizebox{\hsize}{!}{\includegraphics{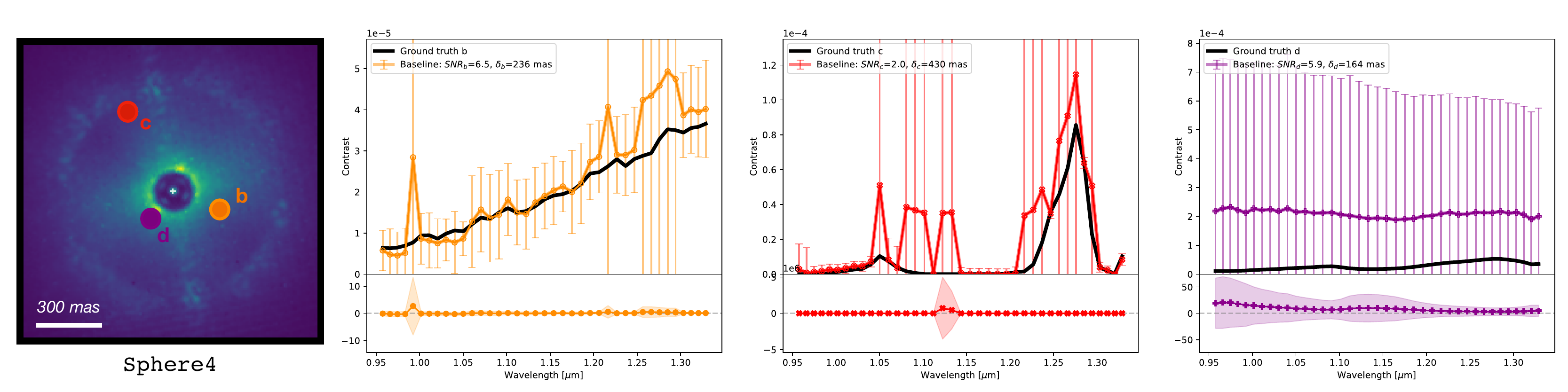}}
    \vspace{0.2cm} 
    \caption{Spectro-photometry: baseline results for each injection in the SPHERE-IFS data sets. From left to right: Temporal  median image at the shortest wavelength with the position of the injected planetary signals (colored circles); Corresponding estimated spectrum with 3$\sigma$ uncertainties (colored lines), compared to the injected spectrum (dark line). The bottom panel shows the residuals between the estimated spectra and the ground-truth (the shaded area corresponds to the 3$\sigma$ uncertainties on the estimations). 
    From top to bottom: data set \textit{sphere1}, \textit{sphere2}, \textit{sphere3}, and \textit{sphere4}.}
    \label{fig:data_spec_sph}
\end{figure}

In this appendix, we first show the results of the astrometric estimations using the baseline method (PCA-SADI+NEGFC) on the four GPI datasets (Fig.~\ref{fig:data_gpi}) and on the four SPHERE-IFS datasets (Fig.~\ref{fig:data_sphere}). 
We then show the  results of the spectro-photometric estimations using the baseline method (PCA-SADI+NEGFC) on the four GPI datasets (Fig.~\ref{fig:data_spec_gpi}) and on the four SPHERE-IFS datasets (Fig.~\ref{fig:data_spec_sph}).

\section{Gallery: Results astrometry}
\label{sec:gallery_all_astr}
\begin{figure}
    \centering
    \resizebox{\hsize}{!}{\includegraphics[trim={0 0.25cm 0 1.25cm},clip]{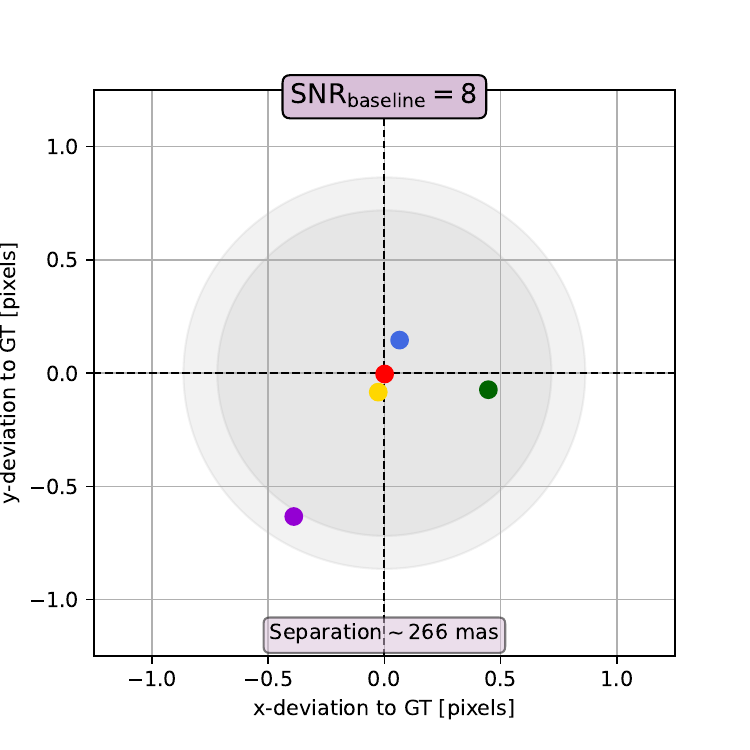}
    \includegraphics[trim={0 0.25cm 0 1.25cm},clip]{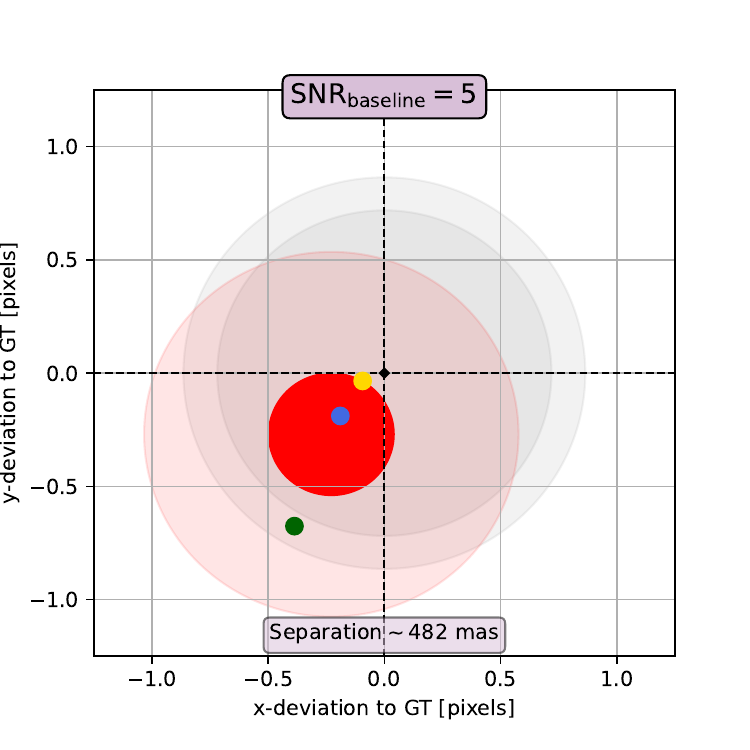}
    \includegraphics[trim={0 0.25cm 0 1.25cm},clip]{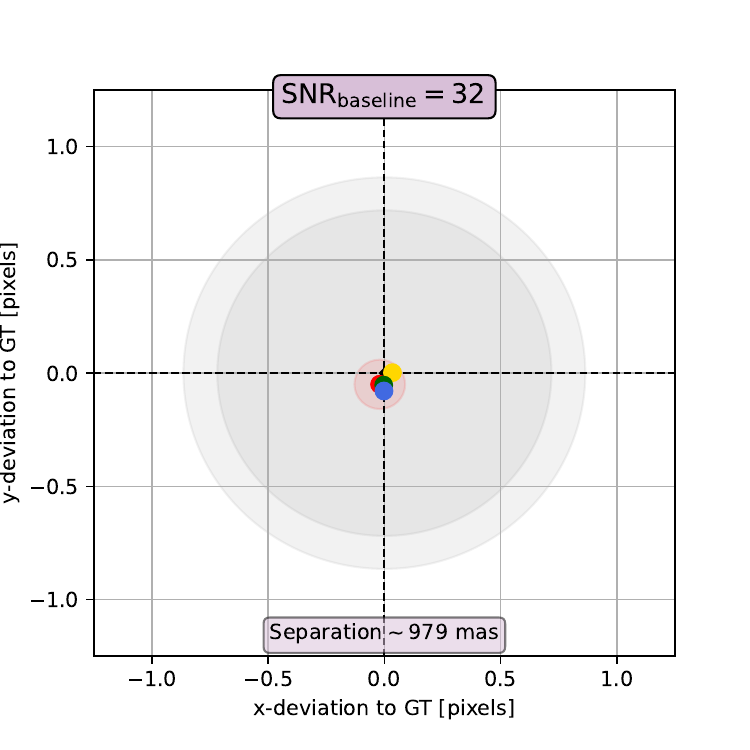}}
    
    \resizebox{\hsize}{!}{\includegraphics[trim={0 0.25cm 0 1.2cm},clip]{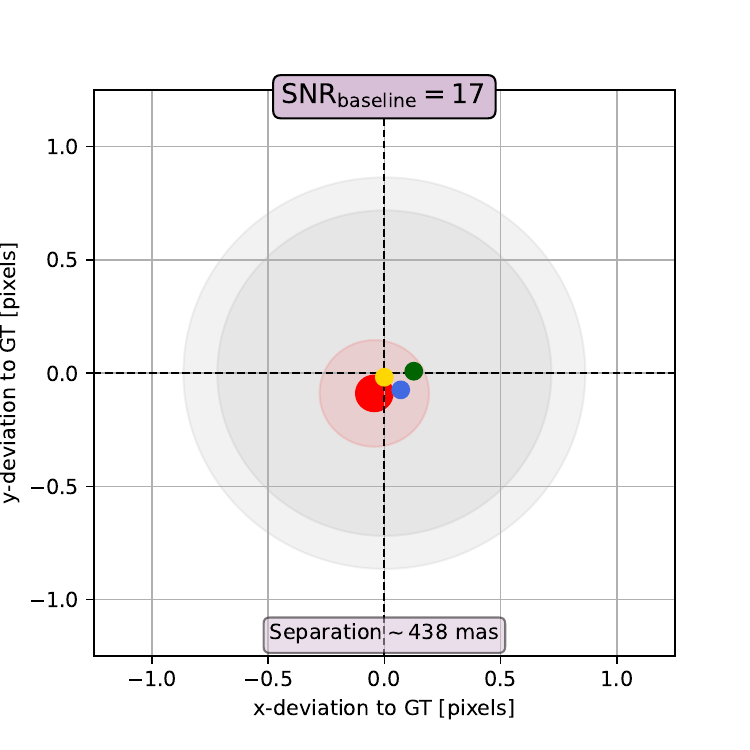}
    \includegraphics[trim={0 0.25cm 0 1.2cm},clip]{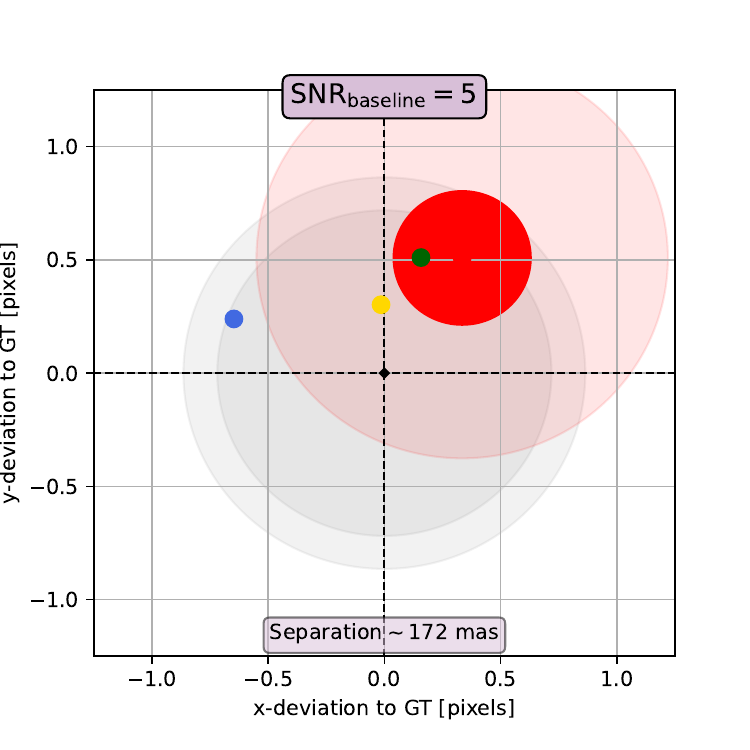}
    \includegraphics[trim={0 0.25cm 0 1.2cm},clip]{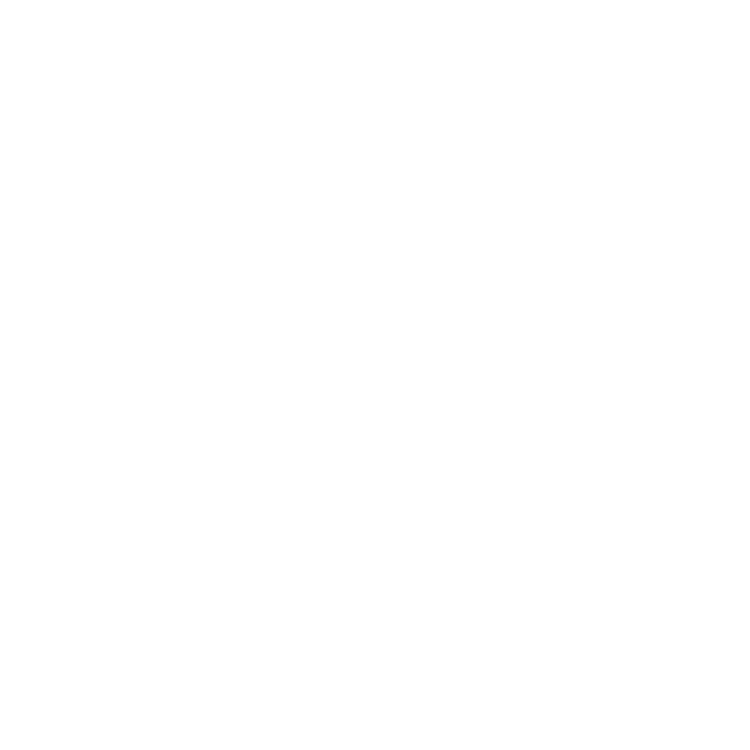}}
    \resizebox{\hsize}{!}{\includegraphics[trim={0 0.25cm 0 1.2cm},clip]{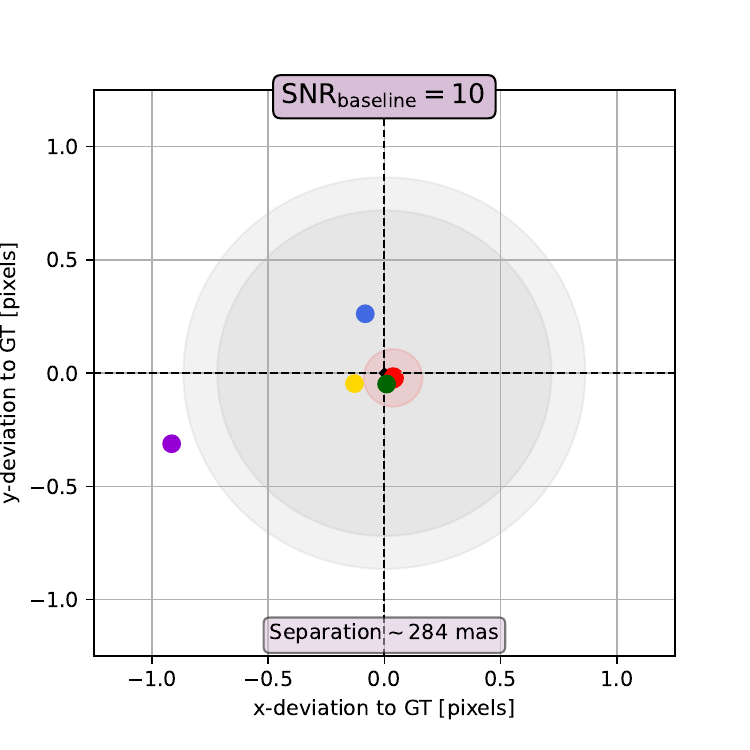}
    \includegraphics[trim={0 0.25cm 0 1.2cm},clip]{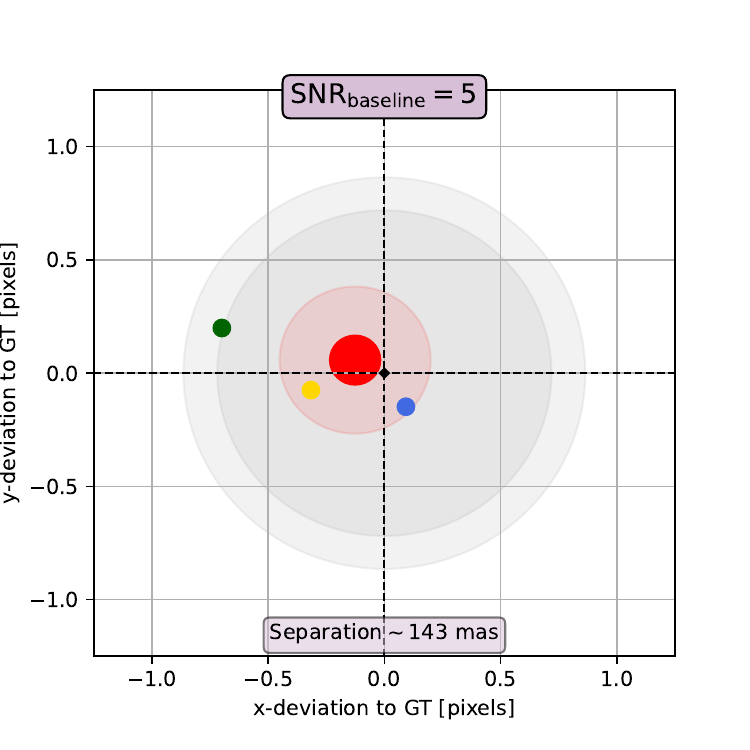}
    \includegraphics[trim={0 0.25cm 0 1.2cm},clip]{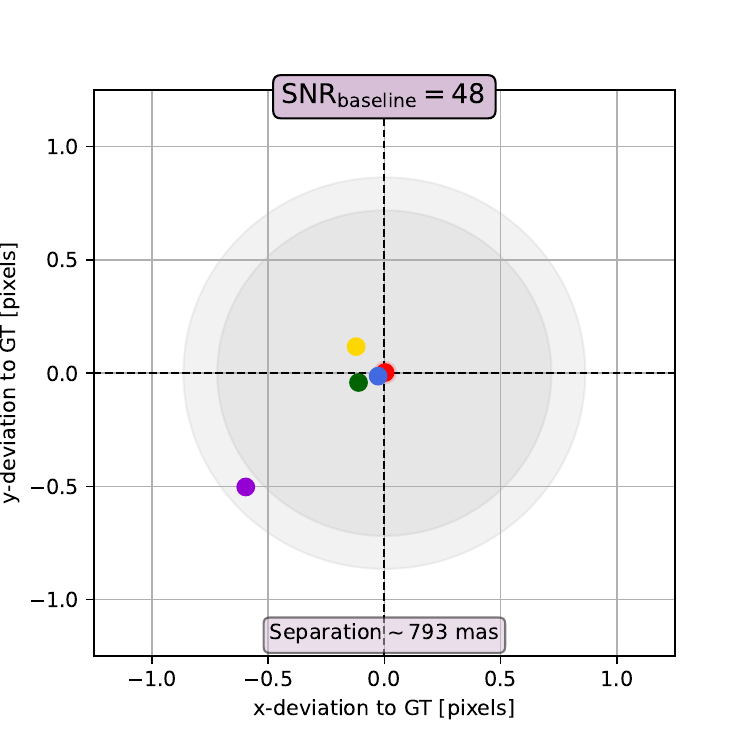}}
    \resizebox{\hsize}{!}{\includegraphics[trim={0 0.25cm 0 1.2cm},clip]{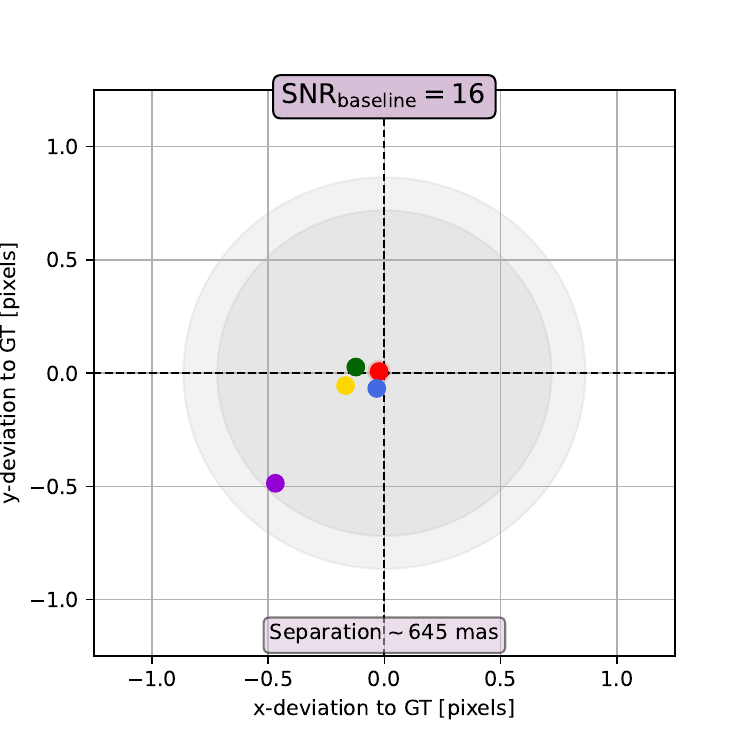}
    \includegraphics[trim={0 0.25cm 0 1.2cm},clip]{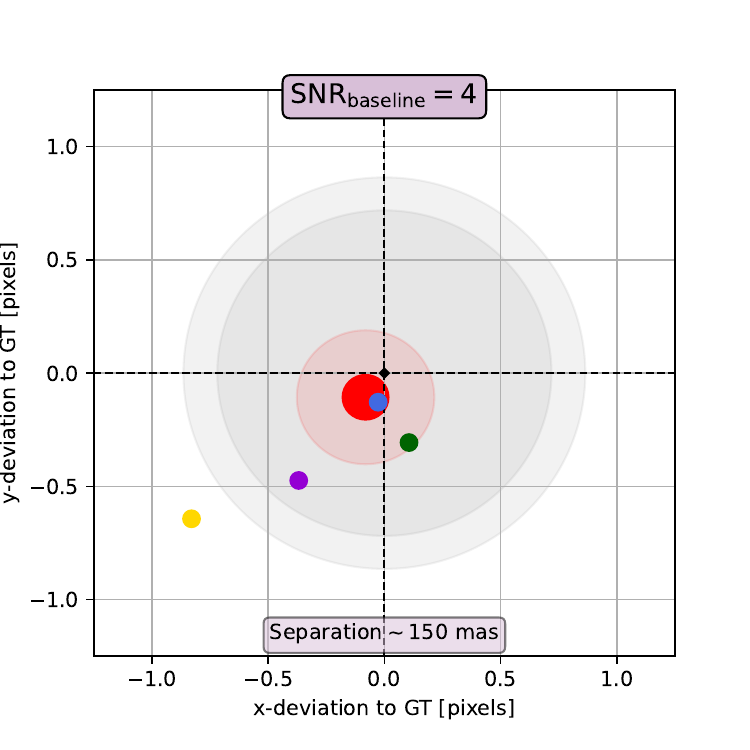}
    \includegraphics[trim={0 0.25cm 0 1.2cm},clip]{Figures_astrometry_2024/Figure_astrometry_blank.pdf}
    }
    \resizebox{\hsize}{!}{\includegraphics{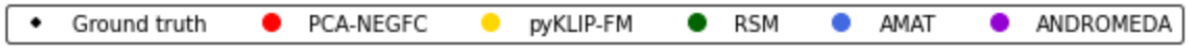}}
    \caption{GPI relative astrometry errors for each injection (in pixels). The ground-truth is shown as a black diamond in the center of the grid. From top to bottom: data set \textit{gpi1}, \textit{gpi2}, \textit{gpi3}, and \textit{gpi4}. The platescale of GPI is $14.16~\mathrm{mas/px}$.}
    \label{fig:astrom_gpi}
\end{figure}

\begin{figure}
    \centering
    \resizebox{\hsize}{!}{\includegraphics[trim={0 0.25cm 0 1.2cm},clip]{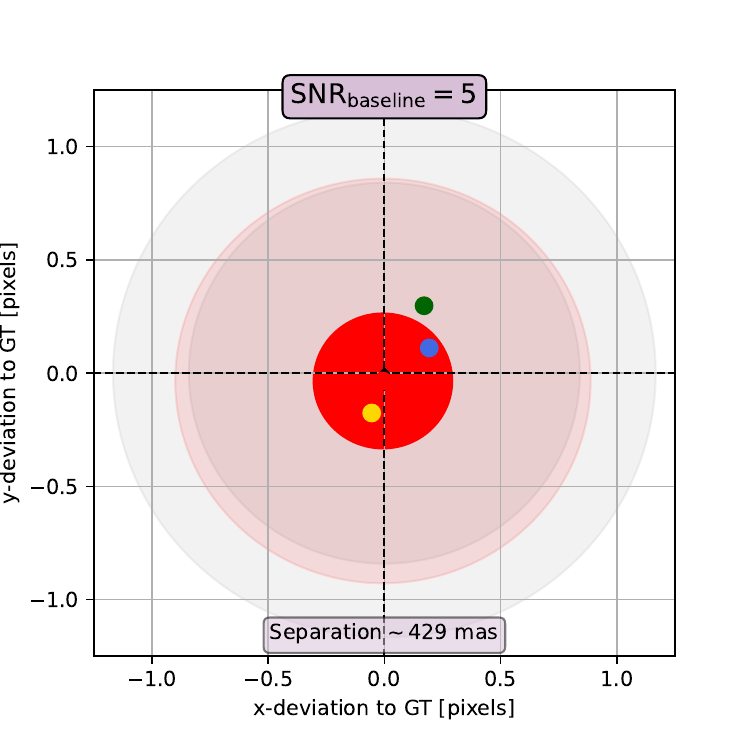}
    \includegraphics[trim={0 0.25cm 0 1.2cm},clip]{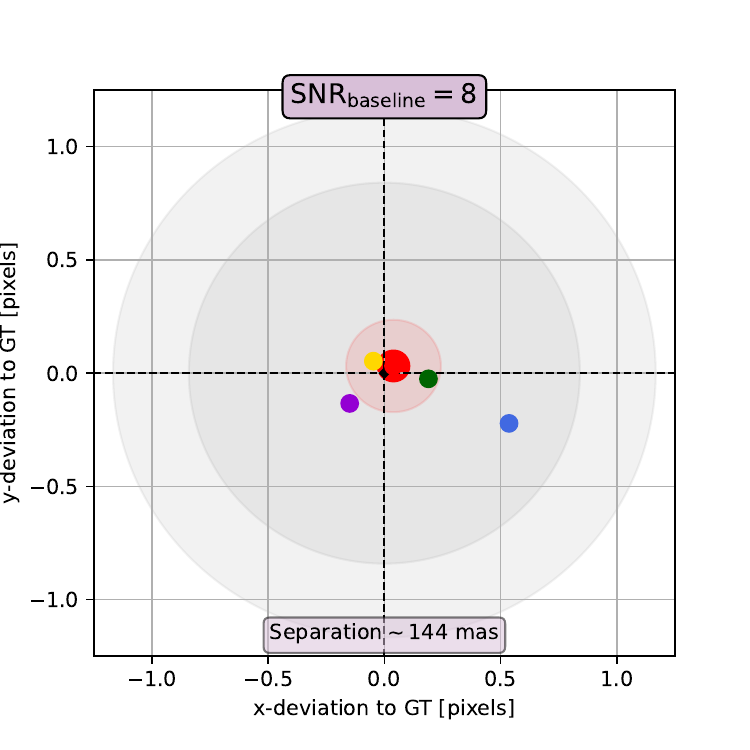}
    \includegraphics[trim={0 0.25cm 0 1.2cm},clip]{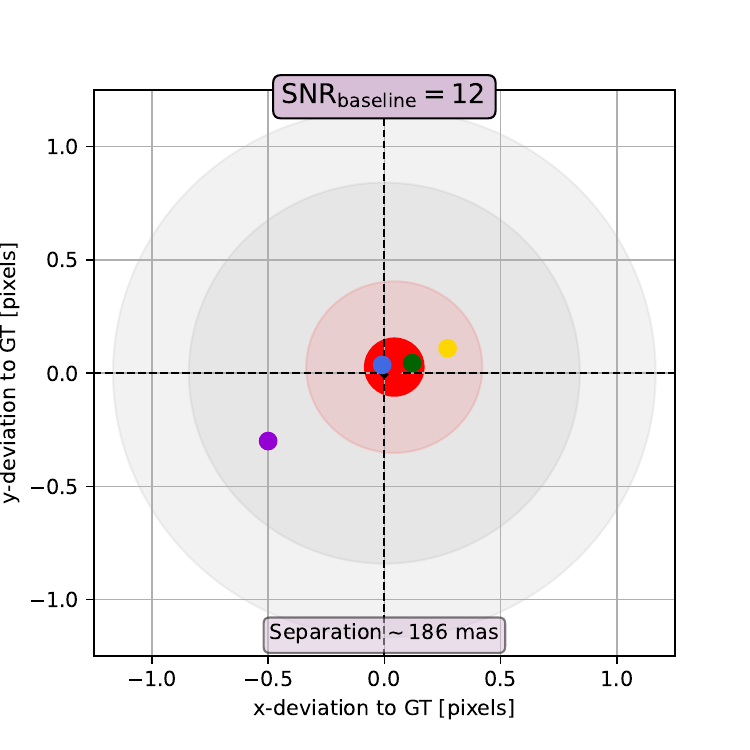}}
    \resizebox{\hsize}{!}{\includegraphics[trim={0 0.25cm 0 1.2cm},clip]{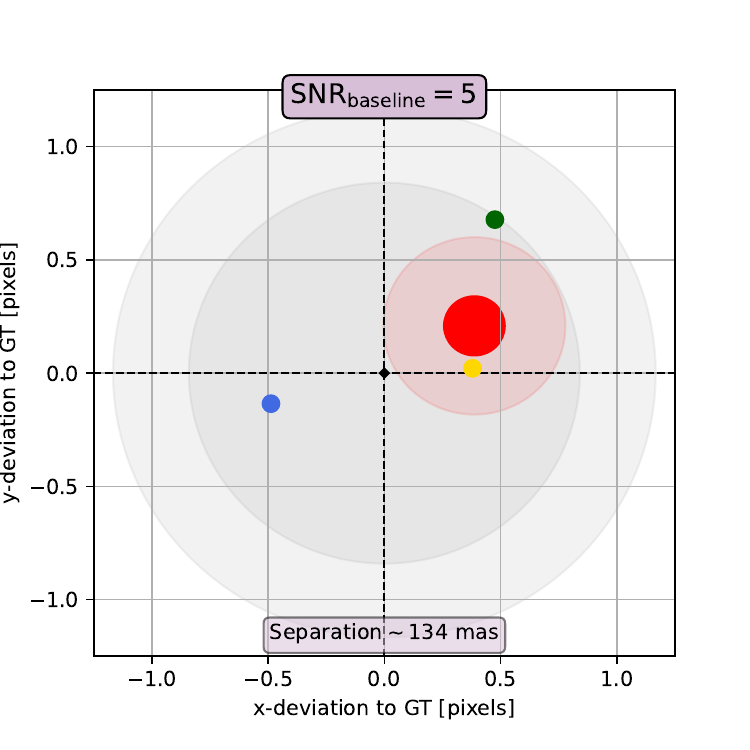}
    \includegraphics[trim={0 0.25cm 0 1.2cm},clip]{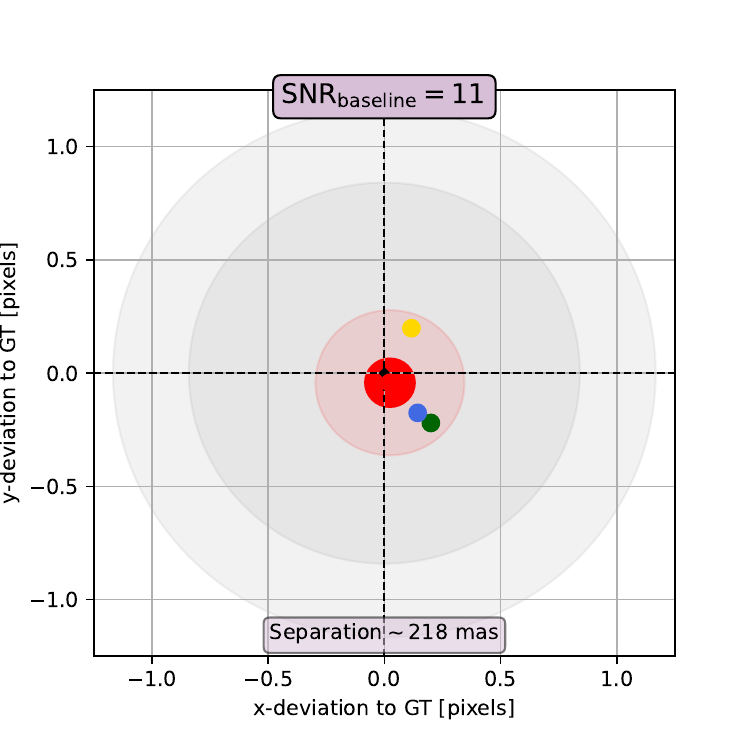}
    \includegraphics[trim={0 0.25cm 0 1.2cm},clip]{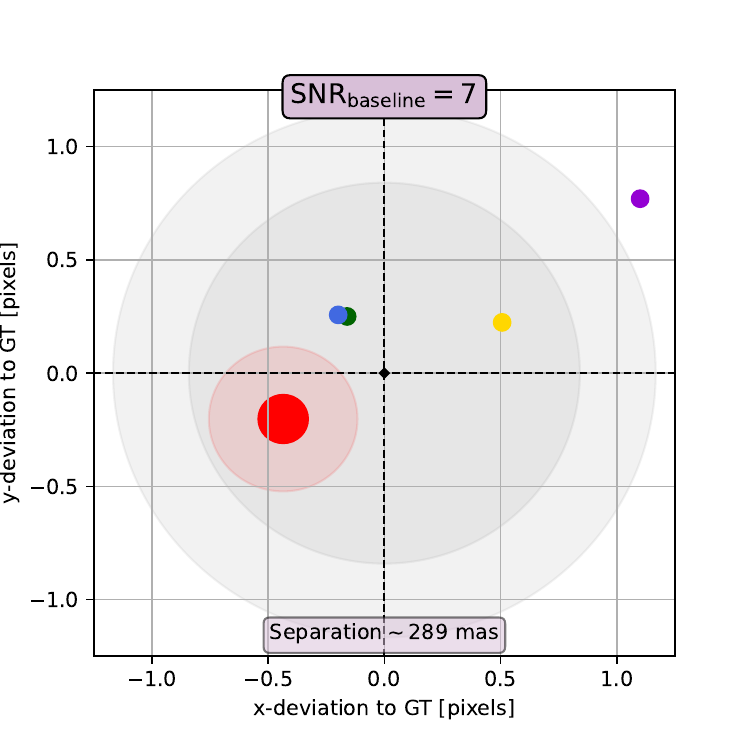}}
    \resizebox{\hsize}{!}{\includegraphics[trim={0 0.25cm 0 1.2cm},clip]{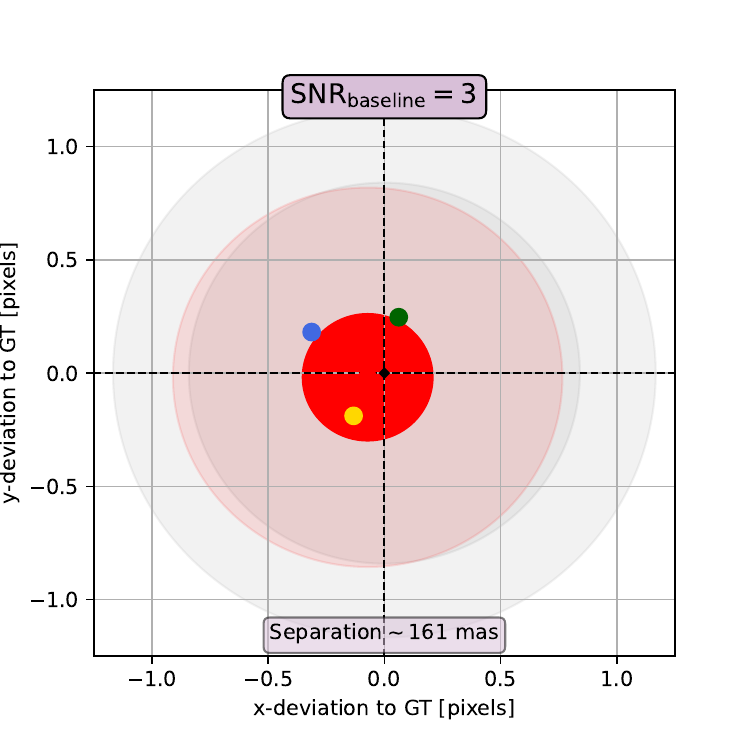}
    \includegraphics[trim={0 0.25cm 0 1.2cm},clip]{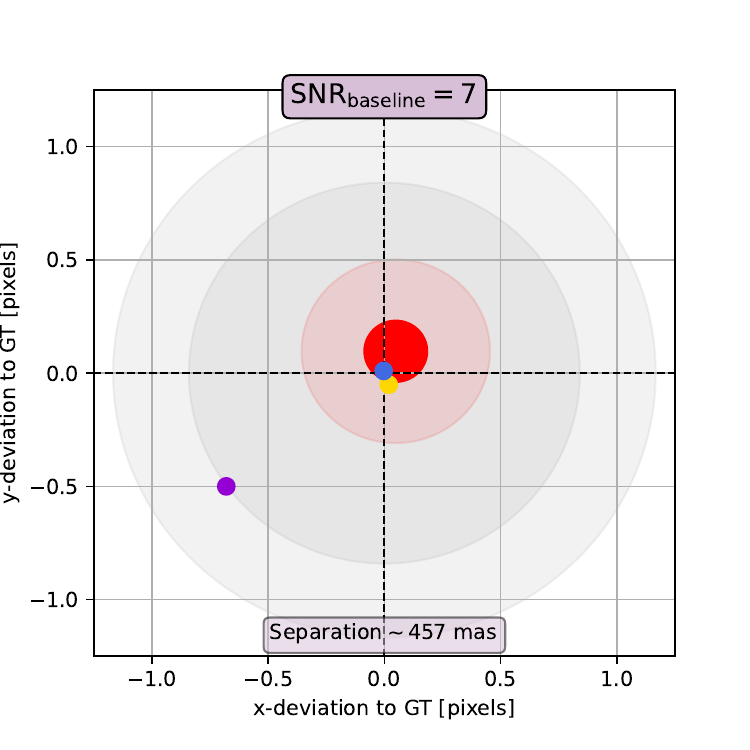}
    \includegraphics[trim={0 0.25cm 0 1.2cm},clip]{Figures_astrometry_2024/Figure_astrometry_blank.pdf}}
    \resizebox{\hsize}{!}{\includegraphics[trim={0 0.25cm 0 1.2cm},clip]{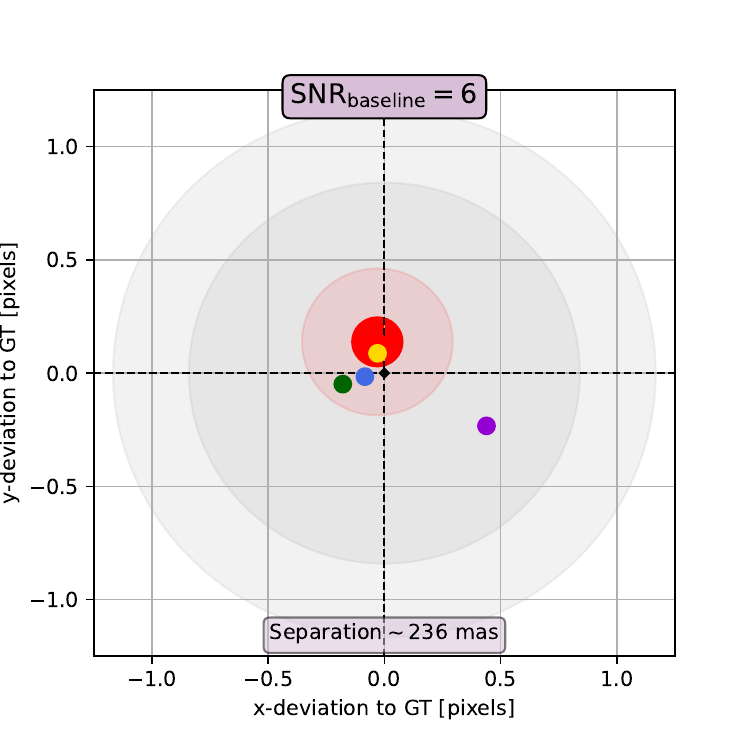}
    \includegraphics[trim={0 0.25cm 0 1.2cm},clip]{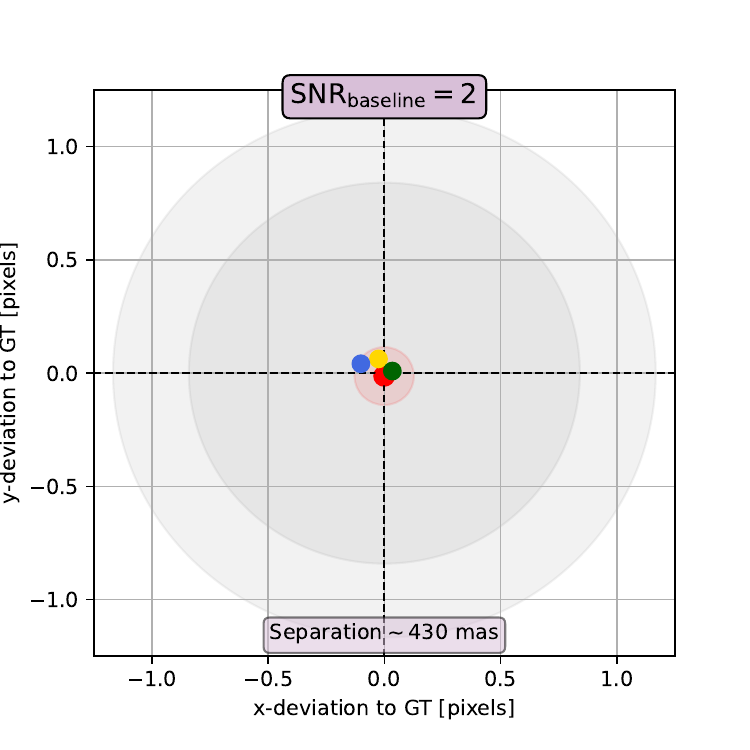}
    \includegraphics[trim={0 0.25cm 0 1.2cm},clip]{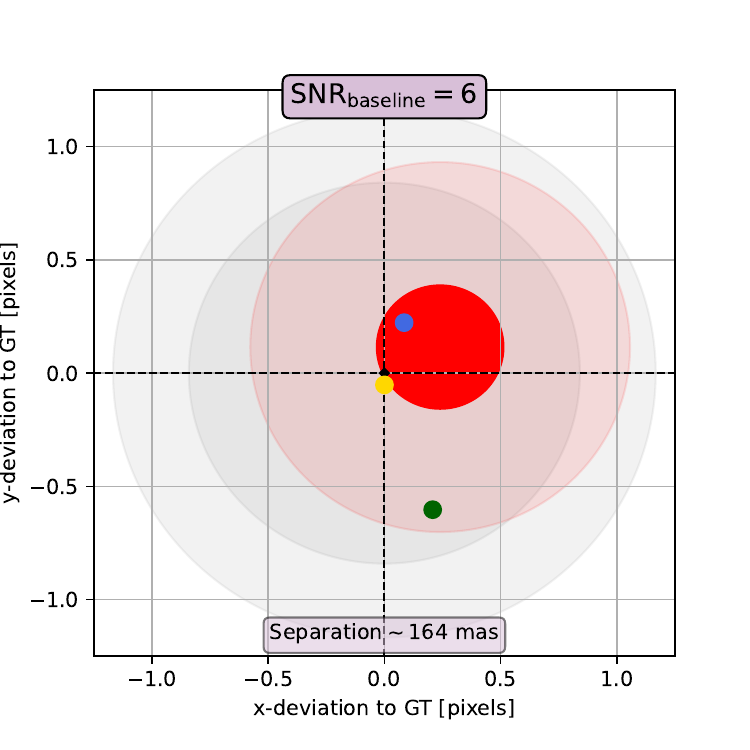}
    }
    \resizebox{\hsize}{!}{\includegraphics{Figures_astrometry_2024/legend.pdf}}
    \caption{SPHERE-IFS relative astrometry errors for each injection (in pixels). The ground-truth is shown at the center. From top to bottom: data set \textit{sphere1}, \textit{sphere2}, \textit{sphere3}, and \textit{sphere4}. The platescale of SPHERE-IFS is $7.46~\mathrm{mas/px}$.}
    \label{fig:astrom_sphere}
\end{figure}

In this appendix, we show the results of the astrometric estimations using the baseline method (red), as well as  the four received results using pyKLIP-FM (gold), RSM (green), AMAT (blue), and ANDROMEDA (purple). 
For each injection, the ground-truth is represented in the middle (black diamond), where the horizontal and vertical black dashed lines intersect. The two centered grey shaded areas represent the size of a half resolution element ($0.5\lambda/D$) at the shortest and largest wavelengths $\lambda$ of the spectro-imager, whose effective diameter is $D$. For the baseline method (in red), the size of the symbol corresponds to the 1-sigma uncertainties (averaged in x- and y- directions) and the shaded area to the 3-sigma uncertainties. 
To ease the reading, we added on top of each plots the SNR of the injected planetary signal obtained with the baseline as well as the separation of the considered injection. 
The four GPI datasets are shown in Fig.~\ref{fig:astrom_gpi} and the four SPHERE-IFS datasets in Fig.~\ref{fig:astrom_sphere}.

\section{Gallery: Results spectro-photometry}
\label{sec:gallery_all_phot}
\begin{figure}
    \centering
    \resizebox{\hsize}{!}{\includegraphics{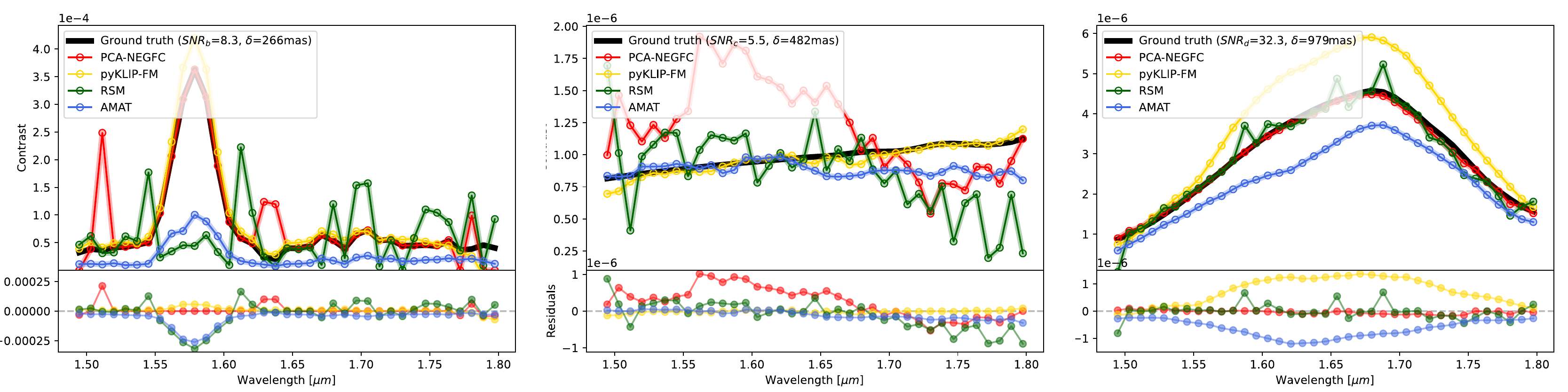}}
    \resizebox{\hsize}{!}{\includegraphics{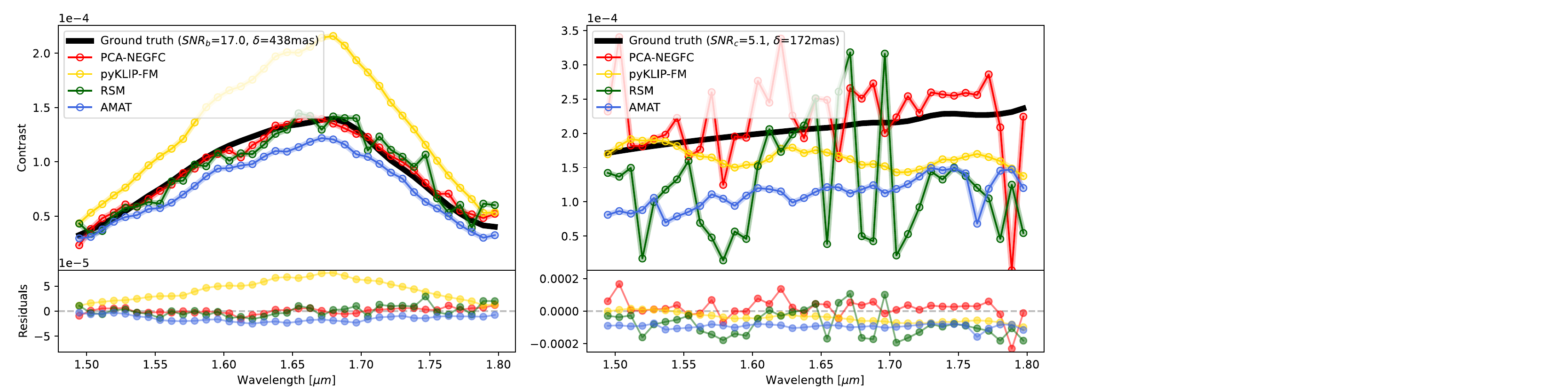}}
    \resizebox{\hsize}{!}{\includegraphics{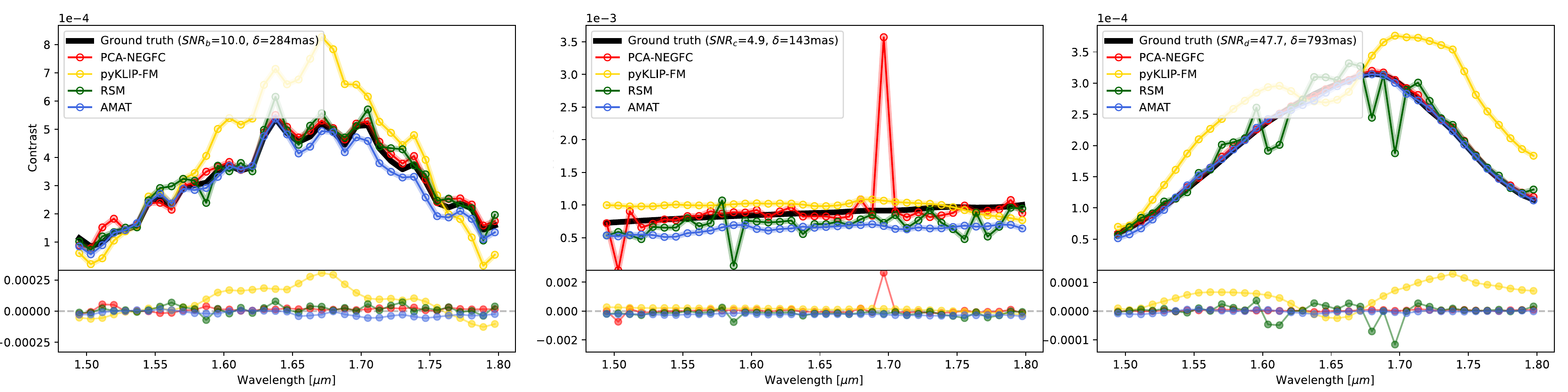}}
    \resizebox{\hsize}{!}{\includegraphics{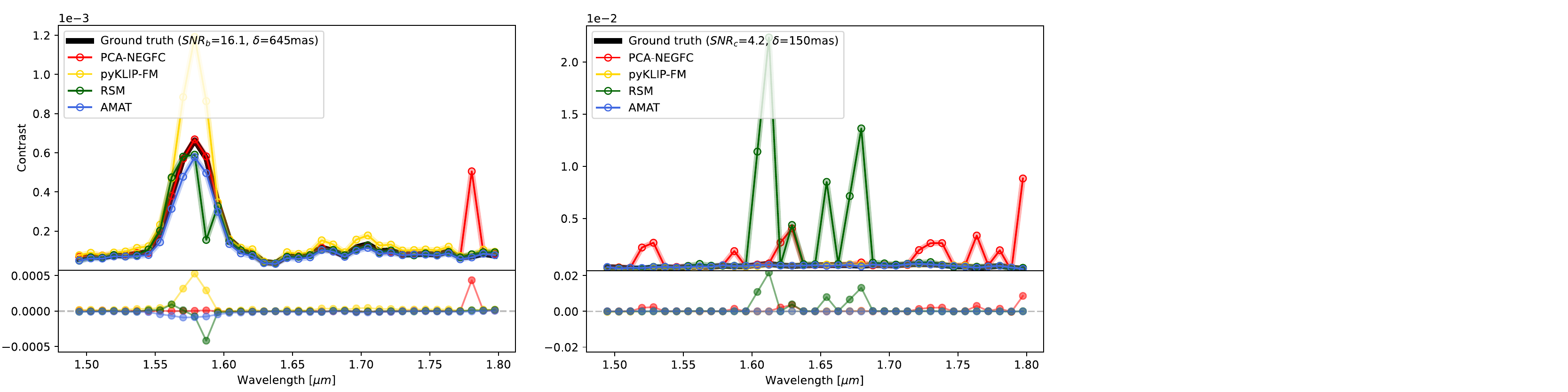}}
    \vspace{0.3cm}
    \caption{GPI estimated spectra for each injected planetary signal (in contrast to the star). The ground-truth is shown as a black solid line. The top panel shows the estimated spectra by each method, while the bottom panel highlights the difference between the ground-truth spectrum and the estimated spectra.
    For reference, the spectra extracted with the baseline method are shown with red solid lines (PCA-SADI with NEGFC). The spectra extracted with the three submissions are shown in gold (pyKLIP-FM), green (RSM), and blue (AMAT).
    From top to bottom: data set \textit{gpi1}, \textit{gpi2}, \textit{gpi3}, and \textit{gpi4}. The wavelength range of GPI is $[1.495-1.797]\mathrm{\mu m}$ (around the H-band), shared in 37 spectral channels.}
    \label{fig:photom_gpi}
\end{figure}

\begin{figure}
    \centering
    \resizebox{\hsize}{!}{\includegraphics{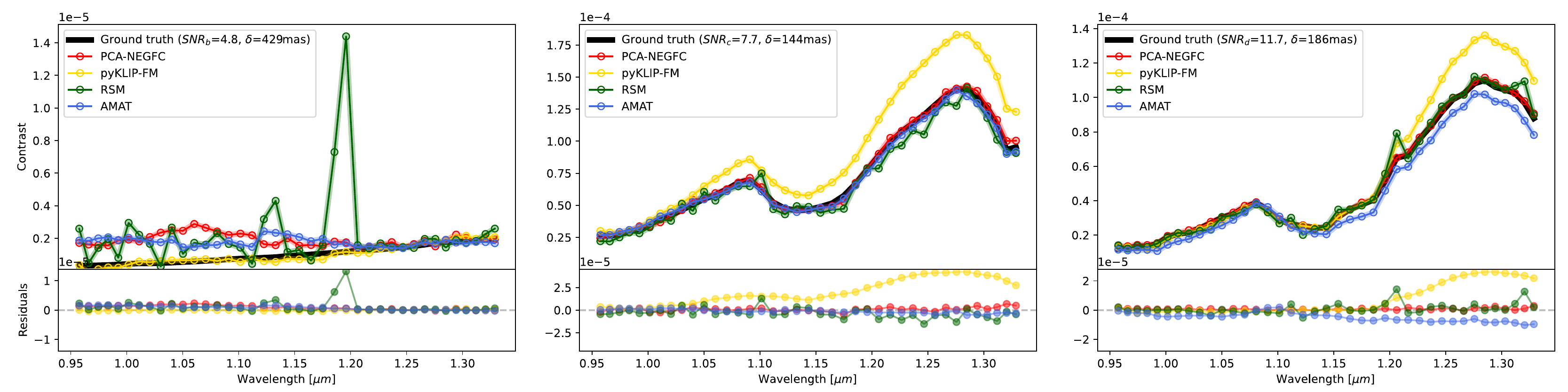}}
    \resizebox{\hsize}{!}{\includegraphics{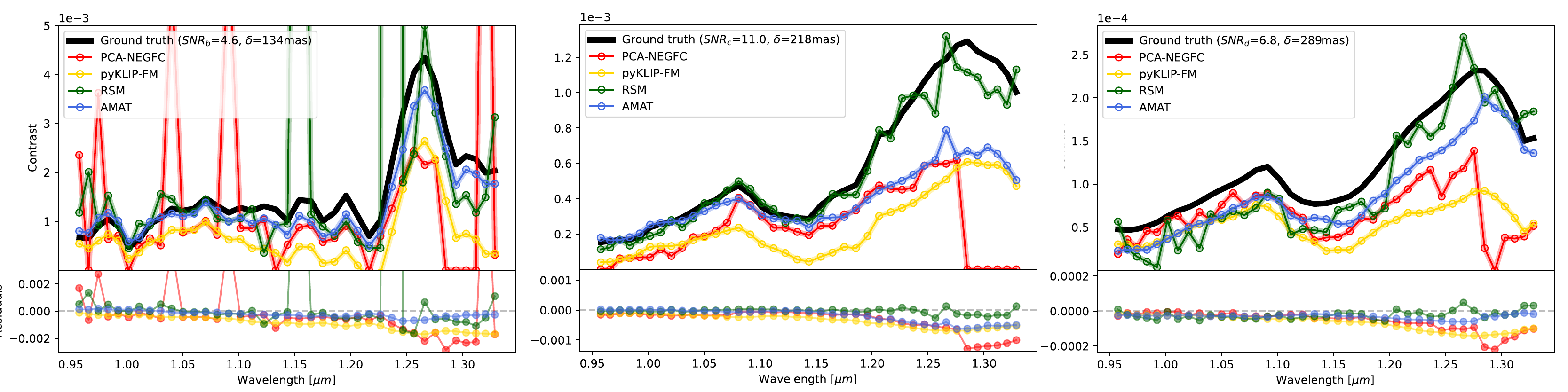}}
    \resizebox{\hsize}{!}{\includegraphics{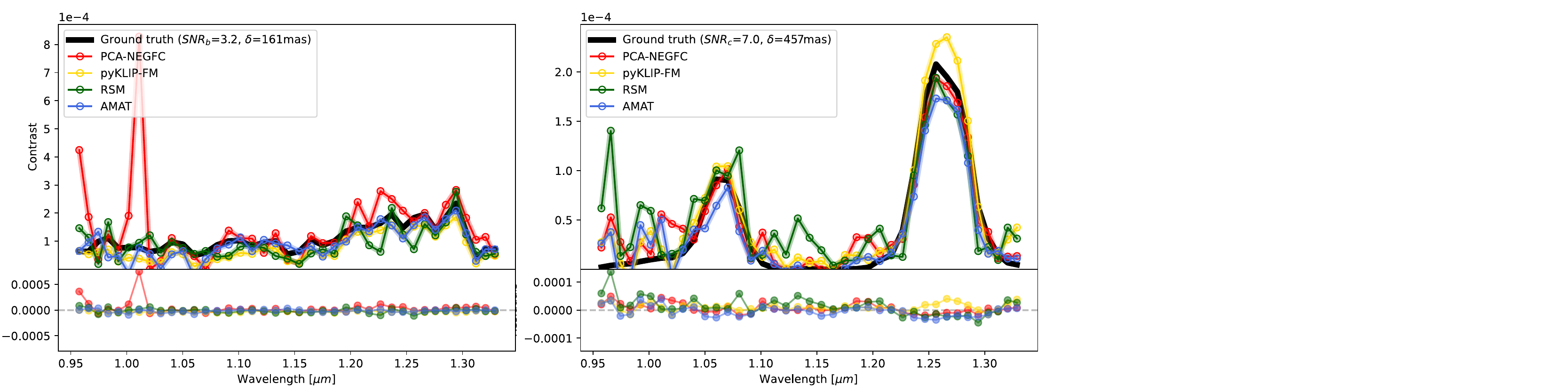}}
    \resizebox{\hsize}{!}{\includegraphics{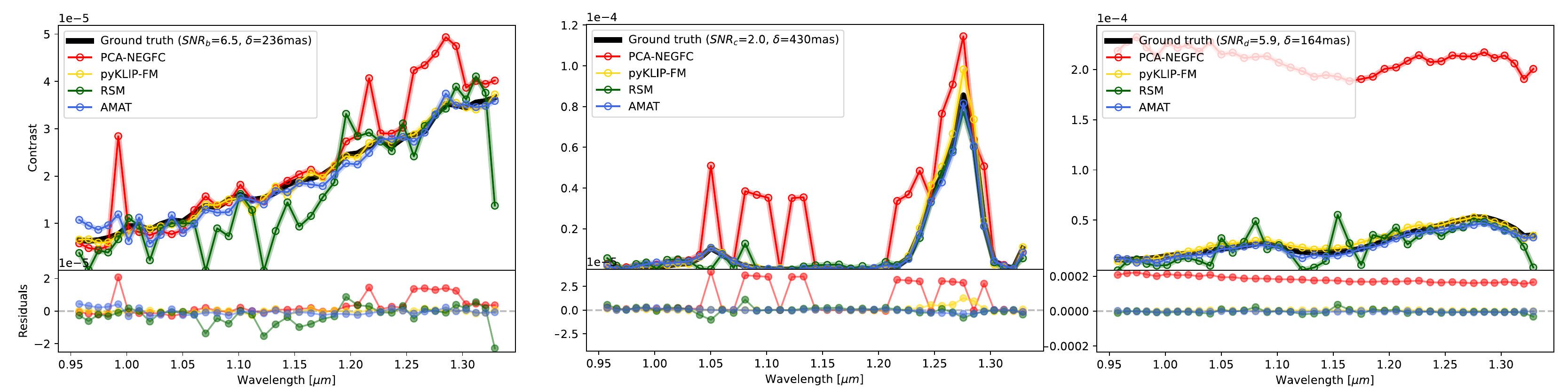}}
    \vspace{0.3cm}
    \caption{SPHERE-IFS estimated spectra for each injected planetary signal (in contrast to the star). The ground-truth is shown as a black solid line. The top panel shows the estimated spectra by each method, while the bottom panel highlights the difference between the ground-truth spectrum and the estimated spectra.
    For reference, the spectra extracted with the baseline method are shown with red solid lines (PCA-SADI with NEGFC). The spectra extracted with the three submissions are shown in gold (pyKLIP-FM), green (RSM), and blue (AMAT). From top to bottom: data set \textit{sphere1}, \textit{sphere2}, \textit{sphere3}, and \textit{sphere4}. The wavelength range of SPHERE-IFS is $[0.957-1.329]\mathrm{\mu m}$ (YJ-bands), shared in 39 spectral channels.}
    \label{fig:photom_sphere}
\end{figure}

In this appendix, we show the results of the spectro-photometric estimations using the baseline method (red), as well as for the three submitted results with  pyKLIP-FM (gold), RSM (green), and AMAT (blue). 
The top panel show the retrieved spectra by each method, with the ground-truth shown as a black solid line. The bottom panel shows the residuals (difference) between the ground-truth and the estimated spectra. 
The spectra corresponding to the four GPI datasets are shown in Fig.~\ref{fig:photom_gpi} and those of the four SPHERE-IFS datasets are shown in Fig.~\ref{fig:photom_sphere}.

\acknowledgments 
This work has been partially supported by the LabEx FOCUS ANR-11-LABX-0013.
This project has received funding from the European Research Council (ERC) under the European Union's Horizon 2020 research and innovation program (grant agreement No 819155), and from the Wallonia-Brussels Federation (grant for Concerted Research Actions). 
VC and OA thank the Belgian F.R.S.-FNRS, and the Belgian Federal Science Policy Office (BELSPO) for the provision of financial support in the framework of the PRODEX Programme of the European Space Agency (ESA) under contract number 4000142531.
This work has made use of the High Contrast Data Centre, jointly operated by OSUG/IPAG (Grenoble), PYTHEAS/LAM/CeSAM (Marseille), OCA/Lagrange (Nice), Observatoire de Paris/LESIA (Paris), and Observatoire de Lyon/CRAL, and supported by a grant from Labex OSUG@2020 (Investissements d’avenir – ANR10 LABX56).

\bibliography{report} 

\begin{thebibliography}{10}

\bibitem{cantalloube2020wdh}
Cantalloube, F., Farley, O.~J., Milli, J., Bharmal, N., Brandner, W., Correia,
  C., Dohlen, K., Henning, T., Osborn, J., Por, E., et~al., ``Wind-driven halo
  in high-contrast images-i. analysis of the focal-plane images of sphere,''
  {\em Astronomy \& Astrophysics}~{\bf 638},  A98 (2020).

\bibitem{sauvage2016}
Sauvage, J.-F., Fusco, T., Lamb, M., Girard, J., Brinkmann, M., Guesalaga, A.,
  Wizinowich, P., O'Neal, J., N'Diaye, M., Vigan, A., et~al., ``Tackling down
  the low wind effect on sphere instrument,'' in [{\em Adaptive Optics Systems
  V}{\nolinebreak\hspace{0.1em}]},   {\bf 9909},  408--416, SPIE (2016).

\bibitem{vigan2022}
Vigan, A., Dohlen, K., N’diaye, M., Cantalloube, F., Girard, J., Milli, J.,
  Sauvage, J.-F., Wahhaj, Z., Zins, G., Beuzit, J.-L., et~al., ``Calibration of
  quasi-static aberrations in exoplanet direct-imaging instruments with a
  zernike phase-mask sensor-iv. temporal stability of non-common path
  aberrations in vlt/sphere,'' {\em Astronomy \& Astrophysics}~{\bf 660},  A140
  (2022).

\bibitem{cantalloube2019}
{Cantalloube}, F., {Dohlen}, K., {Milli}, J., {Brandner}, W., and {Vigan}, A.,
  ``{Peering through SPHERE Images: A Glance at Contrast Limitations},'' {\em
  The Messenger}~{\bf 176},  25--31 (June 2019).

\bibitem{Cantalloube2020eidc}
Cantalloube, F., Gomez-Gonzalez, C., Absil, O., Cantero, C., Bacher, R., Bonse,
  M., Bottom, M., Dahlqvist, C.-H., Desgrange, C., Flasseur, O., et~al.,
  ``Exoplanet imaging data challenge: benchmarking the various image processing
  methods for exoplanet detection,'' in [{\em Adaptive Optics Systems
  VII}{\nolinebreak\hspace{0.1em}]},   {\bf 11448},  1027--1062, SPIE (2020).

\bibitem{eidc2022}
Cantalloube, F., Christiaens, V., Cantero, C., Nasedkin, E., Cioppa, A., Absil,
  O., Bonse, J., Delorme, P., Gomez-Gonzalez, C., Juillard, S., et~al.,
  ``Exoplanet imaging data challenge, phase ii: characterization of exoplanet
  signals in high-contract images,'' in [{\em Adaptive Optics Systems
  VIII}{\nolinebreak\hspace{0.1em}]},   {\bf 12185},  8--24, SPIE (2022).

\bibitem{nielsen2019gpies}
Nielsen, E.~L., De~Rosa, R.~J., Macintosh, B., Wang, J.~J., Ruffio, J.-B.,
  Chiang, E., Marley, M.~S., Saumon, D., Savransky, D., Ammons, S.~M., et~al.,
  ``The gemini planet imager exoplanet survey: giant planet and brown dwarf
  demographics from 10 to 100 au,'' {\em The Astronomical Journal}~{\bf
  158}(1),  13 (2019).

\bibitem{vigan2021shine}
Vigan, A., Fontanive, C., Meyer, M., Biller, B., Bonavita, M., Feldt, M.,
  Desidera, S., Marleau, G.-D., Emsenhuber, A., Galicher, R., et~al., ``The
  sphere infrared survey for exoplanets (shine)-iii. the demographics of young
  giant exoplanets below 300 au with sphere,'' {\em Astronomy \&
  Astrophysics}~{\bf 651},  A72 (2021).

\bibitem{Marois2006}
{Marois}, C., {Lafreni{\`e}re}, D., {Doyon}, R., {Macintosh}, B., and {Nadeau},
  D., ``{Angular Differential Imaging: A Powerful High-Contrast Imaging
  Technique},'' {\em The Astrophysical Journal}~{\bf 641},  556--564 (Apr.
  2006).

\bibitem{lafreniere2007loci}
Lafreniere, D., Marois, C., Doyon, R., Nadeau, D., and Artigau, {\'E}., ``A new
  algorithm for point-spread function subtraction in high-contrast imaging: a
  demonstration with angular differential imaging,'' {\em The Astrophysical
  Journal}~{\bf 660}(1),  770 (2007).

\bibitem{marois2010negfc}
Marois, C., Zuckerman, B., Konopacky, Q.~M., Macintosh, B., and Barman, T.,
  ``Images of a fourth planet orbiting hr 8799,'' {\em Nature}~{\bf 468}(7327),
   1080--1083 (2010).

\bibitem{lagrange2010negfc}
Lagrange, A.-M., Bonnefoy, M., Chauvin, G., Apai, D., Ehrenreich, D.,
  Boccaletti, A., Gratadour, D., Rouan, D., Mouillet, D., Lacour, S., et~al.,
  ``A giant planet imaged in the disk of the young star $\beta$ pictoris,''
  {\em Science}~{\bf 329}(5987),  57--59 (2010).

\bibitem{Mugnier2009}
{Mugnier}, L.~M., {Cornia}, A., {Sauvage}, J.-F., {Rousset}, G., {Fusco}, T.,
  and {V{\'e}drenne}, N., ``{Optimal method for exoplanet detection by angular
  differential imaging},'' {\em Journal of the Optical Society of America
  A}~{\bf 26},  1326 (May 2009).

\bibitem{Pueyo2016}
Pueyo, L., ``Detection and characterization of exoplanets using projections on
  karhunen--loeve eigenimages: Forward modeling,'' {\em The Astrophysical
  Journal}~{\bf 824}(2),  117 (2016).

\bibitem{flasseur2018paco}
Flasseur, O., Denis, L., Thi{\'e}baut, E., and Langlois, M., ``An unsupervised
  patch-based approach for exoplanet detection by direct imaging,'' in [{\em
  2018 25th IEEE international conference on image processing
  (ICIP)}{\nolinebreak\hspace{0.1em}]},   2735--2739, IEEE (2018).

\bibitem{Christiaens2022b}
{Christiaens et al.}, V., ``Special: A python package for the spectral
  characterization of directly imaged low-mass companions,'' {\em subm. to
  JOSS}  (2022).

\bibitem{cantalloube2015andro}
Cantalloube, F., Mouillet, D., Mugnier, L., Milli, J., Absil, O., Gonzalez,
  C.~G., Chauvin, G., Beuzit, J.-L., and Cornia, A., ``Direct exoplanet
  detection and characterization using the andromeda method: Performance on
  vlt/naco data,'' {\em Astronomy \& Astrophysics}~{\bf 582},  A89 (2015).

\bibitem{dahlqvist2020rsm}
Dahlqvist, C.-H., Cantalloube, F., and Absil, O., ``Regime-switching model
  detection map for direct exoplanet detection in adi sequences,'' {\em
  Astronomy \& Astrophysics}~{\bf 633},  A95 (2020).

\bibitem{dahlqvist2021rsmfm}
Dahlqvist, C.-H., Louppe, G., and Absil, O., ``Improving the rsm map exoplanet
  detection algorithm-psf forward modelling and optimal selection of psf
  subtraction techniques,'' {\em Astronomy \& Astrophysics}~{\bf 646},  A49
  (2021).

\bibitem{Soummer2012}
{Soummer}, R., {Pueyo}, L., and {Larkin}, J., ``{Detection and Characterization
  of Exoplanets and Disks Using Projections on Karhunen-Lo{\`e}ve
  Eigenimages},'' {\em The Astrophysical Journal Letter}~{\bf 755},  L28 (Aug.
  2012).

\bibitem{Wangpyklip}
{Wang}, J.~J., {Ruffio}, J.-B., {De Rosa}, R.~J., {Aguilar}, J., {Wolff},
  S.~G., and {Pueyo}, L., ``{pyKLIP: PSF Subtraction for Exoplanets and
  Disks}.'' Astrophysics Source Code Library, record ascl:1506.001 (June 2015).

\bibitem{pueyo2016fm}
Pueyo, L., ``Detection and characterization of exoplanets using projections on
  karhunen--loeve eigenimages: Forward modeling,'' {\em The Astrophysical
  Journal}~{\bf 824}(2),  117 (2016).

\bibitem{daglayan2023amat}
Daglayan~Sevim, H., Vary, S., and Absil, P.-A., ``An alternating minimization
  algorithm with trajectory for direct exoplanet detection,'' in [{\em ESANN
  2023-European Symposium on Artificial Neural Networks, Computational
  Intelligence and Machine Learning}{\nolinebreak\hspace{0.1em}]},  (2023).

\bibitem{cantalloube2022eidc}
Cantalloube, F., Christiaens, V., Cantero, C., Nasedkin, E., Cioppa, A., Absil,
  O., Bonse, J., Delorme, P., Gomez-Gonzalez, C., Juillard, S., et~al.,
  ``Exoplanet imaging data challenge, phase ii: characterization of exoplanet
  signals in high-contrast images,'' in [{\em Adaptive Optics Systems
  VIII}{\nolinebreak\hspace{0.1em}]},   {\bf 12185},  8--24, SPIE (2022).

\bibitem{Macintosh2008}
{Macintosh}, B.~A., {Graham}, J.~R., {Palmer}, D.~W., {Doyon}, R., {Dunn}, J.,
  {Gavel}, D.~T., {Larkin}, J., {Oppenheimer}, B., {Saddlemyer}, L.,
  {Sivaramakrishnan}, A., {Wallace}, J.~K., {Bauman}, B., {Erickson}, D.~A.,
  {Marois}, C., {Poyneer}, L.~A., and {Soummer}, R., ``{The Gemini Planet
  Imager: from science to design to construction},'' in [{\em Adaptive Optics
  Systems}{\nolinebreak\hspace{0.1em}]},  {\em SPIE Proceeding} {\bf 7015},
  701518 (July 2008).

\bibitem{Beuzit2019}
{Beuzit}, J.~L., {Vigan}, A., {Mouillet}, D., {Dohlen}, K., {Gratton}, R.,
  {Boccaletti}, A., {Sauvage}, J.~F., {Schmid}, H.~M., {Langlois}, M., {Petit},
  C., {Baruffolo}, A., {Feldt}, M., {Milli}, J., {Wahhaj}, Z., {Abe}, L.,
  {Anselmi}, U., {Antichi}, J., {Barette}, R., {Baudrand}, J., {Baudoz}, P.,
  {Bazzon}, A., {Bernardi}, P., {Blanchard}, P., {Brast}, R., {Bruno}, P.,
  {Buey}, T., {Carbillet}, M., {Carle}, M., {Cascone}, E., {Chapron}, F.,
  {Charton}, J., {Chauvin}, G., {Claudi}, R., {Costille}, A., {De Caprio}, V.,
  {de Boer}, J., {Delboulb{\'e}}, A., {Desidera}, S., {Dominik}, C., {Downing},
  M., {Dupuis}, O., {Fabron}, C., {Fantinel}, D., {Farisato}, G., {Feautrier},
  P., {Fedrigo}, E., {Fusco}, T., {Gigan}, P., {Ginski}, C., {Girard}, J.,
  {Giro}, E., {Gisler}, D., {Gluck}, L., {Gry}, C., {Henning}, T., {Hubin}, N.,
  {Hugot}, E., {Incorvaia}, S., {Jaquet}, M., {Kasper}, M., {Lagadec}, E.,
  {Lagrange}, A.~M., {Le Coroller}, H., {Le Mignant}, D., {Le Ruyet}, B.,
  {Lessio}, G., {Lizon}, J.~L., {Llored}, M., {Lundin}, L., {Madec}, F.,
  {Magnard}, Y., {Marteaud}, M., {Martinez}, P., {Maurel}, D., {M{\'e}nard},
  F., {Mesa}, D., {M{\"o}ller-Nilsson}, O., {Moulin}, T., {Moutou}, C.,
  {Orign{\'e}}, A., {Parisot}, J., {Pavlov}, A., {Perret}, D., {Pragt}, J.,
  {Puget}, P., {Rabou}, P., {Ramos}, J., {Reess}, J.~M., {Rigal}, F., {Rochat},
  S., {Roelfsema}, R., {Rousset}, G., {Roux}, A., {Saisse}, M., {Salasnich},
  B., {Santambrogio}, E., {Scuderi}, S., {Segransan}, D., {Sevin}, A.,
  {Siebenmorgen}, R., {Soenke}, C., {Stadler}, E., {Suarez}, M., {Tiph{\`e}ne},
  D., {Turatto}, M., {Udry}, S., {Vakili}, F., {Waters}, L.~B.~F.~M., {Weber},
  L., {Wildi}, F., {Zins}, G., and {Zurlo}, A., ``{SPHERE: the exoplanet imager
  for the Very Large Telescope},'' {\em Astronomie \& Astrophysics}~{\bf 631},
  A155 (Nov 2019).

\bibitem{nasedkin2023pp}
Nasedkin, E., Molli{\`e}re, P., Wang, J., Cantalloube, F., Kreidberg, L.,
  Pueyo, L., Stolker, T., and Vigan, A., ``Impacts of high-contrast image
  processing on atmospheric retrievals,'' {\em Astronomy \& Astrophysics}~{\bf
  678},  A41 (2023).

\end{thebibliography}
\bibliographystyle{spiebib} 

\end{document}